\documentclass[aps,prb,twocolumn]{revtex4}
\usepackage{amsfonts}
\usepackage{amsmath}
\usepackage{mathrsfs}
\usepackage{amssymb}
\usepackage{wasysym}
\usepackage{hyperref}
\usepackage[hyphenbreaks]{breakurl}
\usepackage[export]{adjustbox}
\usepackage{subcaption}
\usepackage{extarrows}
\usepackage{cleveref}
\usepackage[normalem]{ulem}
\usepackage{float}% Control the locations of figures [H]
\usepackage{graphicx}
\usepackage{epstopdf}
\usepackage{dsfont}
\usepackage{cases}
\usepackage{rotating}
\usepackage{placeins}
\usepackage[dvipsnames]{xcolor}
\usepackage{placeins}
\usepackage{bm}% bold math
\usepackage[bbgreekl]{mathbbol}
\DeclareMathAlphabet\mathbfcal{OMS}{cmsy}{b}{n}

\newcommand{\Kn}{{\bf  K}}

\newcommand{\rn}{{\bf r}}

\begin{document}
\title{Plasmonic Transverse Dipole Moment in Chiral Fermion Nanowires}
%\author{}
%\affiliation{}
\author{Jinlyu Cao$^{1,2}$, H.A.Fertig$^{1,2}$, and Luis Brey$^3$}
\affiliation{
$^1$ Department of Physics, Indiana University, Bloomington, IN 47405\\
$^2$ Quantum Science and Engineering Center, Indiana University, Bloomington, IN, 47408\\
$^3$ Instituto de Ciencia de Materiales de Madrid, (CSIC),
Cantoblanco, 28049 Madrid, Spain\\
}
\date{\today}
\begin{abstract}
Plasmons are elementary quantum excitations of conducting materials with Fermi surfaces.  In two dimensions they may carry a static dipole moment that is transverse to their momentum which is quantum geometric in nature, the quantum geometric dipole (QGD).  We show that this property is also realized for such materials confined in nanowire geometries. Focusing on the gapless, intra-subband plasmon excitations, we compute the transverse dipole moment $\mathcal{D}_x$ of the modes for a variety of situations.  We find that single chiral fermions generically host non-vanishing $\mathcal{D}_x$, even when there is no intrinsic gap in the two-dimensional spectrum, for which the corresponding two-dimensional QGD vanishes.  {In the limit of very wide wires, the transverse dipole moment of the highest velocity plasmon mode matches onto the two-dimensional QGD.} Plasmons of multi-valley systems that are time-reversal symmetric have vanishing transverse dipole moment, but can be made to carry non-vanishing values by breaking the valley symmetry, for example via magnetic field.  The presence of a non-vanishing transverse dipole moment for nanowire plasmons in principle offers the possibility of continuously controlling their energies and velocities by the application of a static transverse electric field.
\end{abstract}
\maketitle

\section{Introduction}
Plasmons are fundamental excitations of metals, in which electronic charge oscillates against the fixed positively charged background of a material, with accompanying electric fields that allow for self-sustaining collective motion \cite{Pines-Nozieres,Bohm:1953aa,Vignale:book,Kittel_2004}.  The behavior can be understood at a semiclassical level, by solving Maxwell's equations in the presence of a frequency-dependent conductivity, which encodes information about the electron dynamics in the material \cite{Dressel:2002aa}.  Beyond their bulk realization,
plasmons may be confined to the surfaces of some solids, with charge oscillations whose amplitudes evanesce quickly inside the material \cite{Ritchie_1957,Pitarke_2006}.  The development of two-dimensional electron systems in semiconductors \cite{Ando_1982} allowed such confined plasmons \cite{Stern_1967,DasSarma_1984} to be realized with a degree of tunability not possible at the surface of a bulk material.  In more recent years, the advent of van der Waals materials, particularly graphene, has greatly enriched the set of interesting physical possibilities for  two-dimensional plasmons \cite{Wunsch_2006,Hwang_2007,Grigorenko:2012aa,Luo_2013,Yu_2015,Goncalves:2016aa}.
These include a myriad of applications and phenomena,
%\cite{2010Sarid,2006Shalaev},
in areas as diverse as terahertz radiation,
%\cite{2011Ju}
biosensing,
%\cite{1999Xu}
photodetection,
%\cite{2011Knight}
quantum computing
%\cite{2019Calafell},
and more \cite{Hutter_2004,
Sekhon_2011,Nikitin:2011aa,
Thygesen_2017,Agarwal_2018,Linic_2011,Ju_2011,Suarez-Morell:2017aa,
Stauber_2020,Woessner_2015,Alcaraz_2018,Giri_2020,Ni_2015,Brey:2020aa}.

Beyond all this, plasmons are interesting for basic physical reasons: they represent quantum, bosonic excitations of charged fermions with a Fermi
surface \cite{Sawada:1957aa,Gell-Mann:1957aa}. Their quantum nature can in principle become evident through manifestations of their quantum geometry.  In two-dimensional materials this nature becomes particularly important because it makes possible strong light-matter interactions, allowing for probes well below the wavelength of light at plasmonic frequencies \cite{Tame_2013,Fitzgerald_2016,Bozhevolnyi_2017,Zhou_2019}. Moreover, Berry curvature in the electronic structure of the host material may lead to chiral behavior even in the absence of a magnetic field \cite{Song_2016}. More generally quantum effects may lead to non-reciprocal behavior of plasmons
%\cite{Sabbaghi_2015,Duppen_2016,Shi_2018,Papaj_2020}.
\cite{Shi_2018,Papaj_2020}. Indeed, in some systems plasmons have internal structure in the form of a static dipole moment, which leads to non-reciprocity in their scattering from point impurities or other circularly symmetric scattering centers \cite{2021PlasmonQGD}.  This quantum geometric dipole (QGD) moment is present in collective excitations of insulators -- excitons -- as well \cite{Cao_2021}.  In both cases, the QGD is transverse to the momentum of the collective mode.

An interesting question is whether effects of this transverse dipole moment can be directly observed, independent of its impact on scattering.  One way to approach this question is to consider its effect on plasmons in a confined geometry, which may tend to orient the dipole moment in a way that allows coupling to electric fields.  The simplest such geometry is quasi-one-dimensional, in which one might expect the dipole moment to align perpendicular to the system cross-section.  This is the subject of our study.  In what follows, we consider a system that supports a plasmonic QGD, a layer of gapped chiral fermions, in which the single-particle states are confined to be within a narrow channel.  Such systems arise naturally in the context of transition metal dichalcogenides (TMD's) \cite{Xiao_2012} and for graphene, {which, when placed on a boron nitride or silicon carbide substrate, may develop a gap at its Dirac points as large as 0.5eV \cite{Nevius:2015aa,Jariwala:2011}}
% for which a substrate may create a gap at the Dirac point. (WHAT REF?)
The single-particle electronic structure of such nanowires is sensitive to the precise nature of their edges \cite{Bollinger_2001,Brey_2006,Brey_2006b,2008Akhmerov}, and may or may not involve the mixing of valleys \cite{Brey_2006,Palacios_2010}.  For simplicity, our studies focus on infinite mass boundary conditions \cite{1987Berry} for which there is no such valley mixing.

Plasmons in nanowires of chiral fermions \cite{2007GrapheneRibbon} share many properties with those of scalar fermions \cite{1989DasSarma,Hu_1990}.  Prominent among these are the presence of collective modes with an essentially linear dispersion, and gapped intersubband modes.
%As we explain below, our results demonstrate some interesting differences.  Most prominent is the fact that the confined chiral fermion system generically hosts a collection of gapless plasmon modes above the particle-hole continuum of the non-interacting system, one for each electric subband that is occupied in the ground state.  This contrasts with the situation in more conventional materials, for which there is a single gapless plasmon mode.
%Fig. XXXXX illustrates an example of this.
As we explain below,
for a single chiral fermion, all these modes support transverse dipole moments with magnitude proportional to the longitudinal momentum of the plasmon.  This orthogonality of the dipole moment and momentum is exactly as one finds for the QGD in two dimensions \cite{2021excitonQGD,2021PlasmonQGD}.  Interestingly, in these quasi-one dimensional systems it appears even when there is no intrinsic gap in the spectrum, so that the QGD vanishes in the corresponding two dimensional system \cite{2021excitonQGD,2021PlasmonQGD}, the gaps introduced by the transverse confinement stabilize the dipole moment.  For systems with pairs of Dirac points connected by time reversal symmetry, the transverse dipole moment vanishes, but non-vanishing values can be introduced into their plasmons by breaking this symmetry, for example with a magnetic field.  An interesting physical consequence of this physics is that this dipole moment can be coupled to a transverse electric field \cite{Pizzochero_2021}, allowing a degree of continuous control over the plasmon frequency and velocity that is unavailable in other nanowire systems.

This article is organized as follows.  In Section II we discuss the single-particle wavefunctions for the confined states, and, when appropriate, for edge states of the massive chiral fermions we consider, assuming infinite mass boundary conditions.  Section III is devoted to a discussion of how we derive plasmon spectra and the dipole moments associated with them.  This is followed by a description of our numerical results in Section IV.  In Section V contains a summary and discussion of our results.  Our paper also contains three appendices.  Appendix A presents further details of how the single particle states are derived.  In Appendix B demonstrate that there are multiple gapless plasmon modes in the systems we consider, focusing on the case of a system with two occupied transverse states as an example.  Finally, Appendix C presents a
an explicit expression for the plasmon transverse dipole moment that is specifically appropriate for infinite mass boundary conditions.

\section{Chiral Fermions on a Nanowire: Single Particle States}
%We begin with the eigen state wave function that satisfy the infinite mass boundary condition for the ribbon placed along the $y$ direction, i.e. one side at $x=0$ and the other at $x=L$ (the ribbon width).
We begin by deriving the single-particle states for our nanowire models.  The Hamiltonian we adopt for the non-interacting system in two dimensions is
\begin{equation}\label{eq: Dirac Material H}
  H^{\tau}_0 =
  \left(
    \begin{array}{cc}
      m & -i\tau \partial_x - \partial_y \\
      -i\tau \partial_x + \partial_y & -m \\
    \end{array}
  \right),
\end{equation}
where $\tau=\pm 1$ indicates a valley degree of freedom, here in TMD materials $\tau=1(-1)$ corresponds to the $K$($K'$) valley. We have adopted units such that $\hbar=v_F=1$, where $v_F$ is the velocity of the chiral particles in the absence of the mass parameter $m$. Its spectrum has a gap $\Delta = 2m$.  This Hamiltonian is an appropriate long-wavelength description of TMD materials when excitations involving spin flips are ignored, and in the limit $m \rightarrow 0$ it also describes the single particle physics of graphene \cite{Castro_Neto_2009,Xiao_2012,Katsnelson:2012aa}.
Eigenstates of $H_0$ consist of right-and left-moving solutions in the $\hat x$-direction,
$$
  \left(
    \begin{array}{c}
      E+m \\
      \pm\tau k_x + i k_y \\
    \end{array}
  \right) e^{\pm i k_x x+ i k_y y}
$$
with energies
 \begin{equation}\label{eq: ribbon state energy}
  E = \sqrt{m^2+ k_x^2+ k_y^2}.
\end{equation}
We choose an orientation in which the electrons are confined in the $\hat{x}$ direction and are free to move along $\hat{y}$.  To confine the electrons, we adopt for simplicity infinite mass boundary conditions \cite{1987Berry},
\begin{equation}\label{eq: infinte mass BC ribbon_main}
  \left.\frac{\psi_1}{\psi_2}\right|_{x=0,L} = \pm i\lambda_0,
\end{equation}
where $\lambda_0  = \text{sgn} (C)$ is the sign of the Chern number outside the wire, and in writing this we have assumed $m \ge 0$. (Details of how one arrives at Eq. \ref{eq: infinte mass BC ribbon_main} are presented in Appendix A.) Note that, without loss of generality, we may assume a Chern number (of magnitude $1/2$ \cite{Girvin_2019}) for each valley in the material (i.e., inside the wire) with the sign of $\lambda$ given by $\tau$.  This choice of boundary condition has the advantage of admitting confined solutions without admixing valleys, but the resulting confined states depend on their momentum $k_y$ along the wire. This latter property is generic for chiral fermions \cite{Brey_2006,2007GrapheneRibbon,2008Akhmerov}, although in the special case
 where
 there are only two valleys, with equal momentum components along the wire direction, this momentum dependence is lifted \cite{Brey_2006} (at the cost of admixing valleys.)  The momentum dependence of the confined wavefunctions is a significant difference from the typical situation for electrons with scalar single-particle states \cite{1989DasSarma}.  Note that the parameter $\lambda_0$ enters the boundary condition because one must choose the mass term outside the wire to tend to either $\infty$ or $-\infty$, and the choice of this sign determines whether the wire supports edge states, as we discuss further below.

Eigenstates of $H_0$ which satisfy the boundary conditions have the form (see Appendix A)
\begin{widetext}
\begin{equation}\label{eq: ribbon state}
  \vec{\psi}_{\vec{k}}^{\tau}(\vec{r}) =
   A(\tau; \vec{k})
  \left(
    \begin{array}{c}
      E+m \\
      \tau k_x + i k_y \\
    \end{array}
  \right) e^{i k_x x + i k_y y}
  + B(\tau; \vec{k})
  \left(
    \begin{array}{c}
      E+m \\
      -\tau k_x + i k_y \\
    \end{array}
  \right)e^{- i k_x x + i k_y y},
\end{equation}
with
\begin{equation}\label{eq: A ribbon state(main text)}
  A(\tau; \vec{k}) = \tau N \sqrt{(E+m)^2+ (\tau k_x-i k_y)^2} \sqrt{m + i \lambda_0 \tau k_x},
\end{equation}
\begin{equation}\label{eq: B ribbon state(main text)}
  B(\tau; \vec{k}) = -\tau N \sqrt{(E+m)^2+ (\tau k_x+i k_y)^2} \sqrt{m - i \lambda_0 \tau k_x},
\end{equation}
with a normalization constant given by
\begin{equation}\label{eq: ribbon N}
  N =\left\{ 8 L_y E(E+m)^2 \left[ L (m^2+ k_x^2)+  \lambda_0 \tau m  \right]\right\}^{-1/2},
\end{equation}
\end{widetext}
in which $L_y$ and $L$ are the length and the width  of the wire respectively.
%{\bf I believe L was not defined}
The allowed values of $k_x$ satisfy the transcendental equation
\begin{equation}\label{eq: bulk state kx condition original form}
  e^{2i k_x L} = \frac{m - i \lambda_0 \tau k_x}{m + i \lambda_0 \tau k_x},
\end{equation}
which in turn quantizes their values,
\begin{equation}\label{eq: bulk state kx quantization}
  k_x L =- \lambda_0 \tau  \arctan (\frac{k_x}{m})+ n \pi,
\end{equation}
where $n = 1,2,3,...$

In addition to these confined states, there may also be edge states,
depending on the relative sign of the wire and the vacuum Chern numbers. In systems with time reversal symmetry, these come in pairs on each edge running in opposite directions, with the member of each pair associated with one or the other valley.  For systems with a single chiral fermion, for which time reversal symmetry is necessarily broken, a single edge state is present on each edge.
These edge states exist only when the wire material and vacuum are topologically distinct, i.e.
the signs of their Chern numbers are opposite,
\begin{equation}\label{eq: opposite Chern number sign}
  \lambda_0 \lambda = -1.
\end{equation}

\begin{figure*}
    \centering
    \begin{subfigure}[b]{0.475\textwidth}
        \caption[]{}%
        %{{\small Single valley subbands $\lambda \lambda_0=-1$}}
        \centering
        \includegraphics[width=\textwidth]{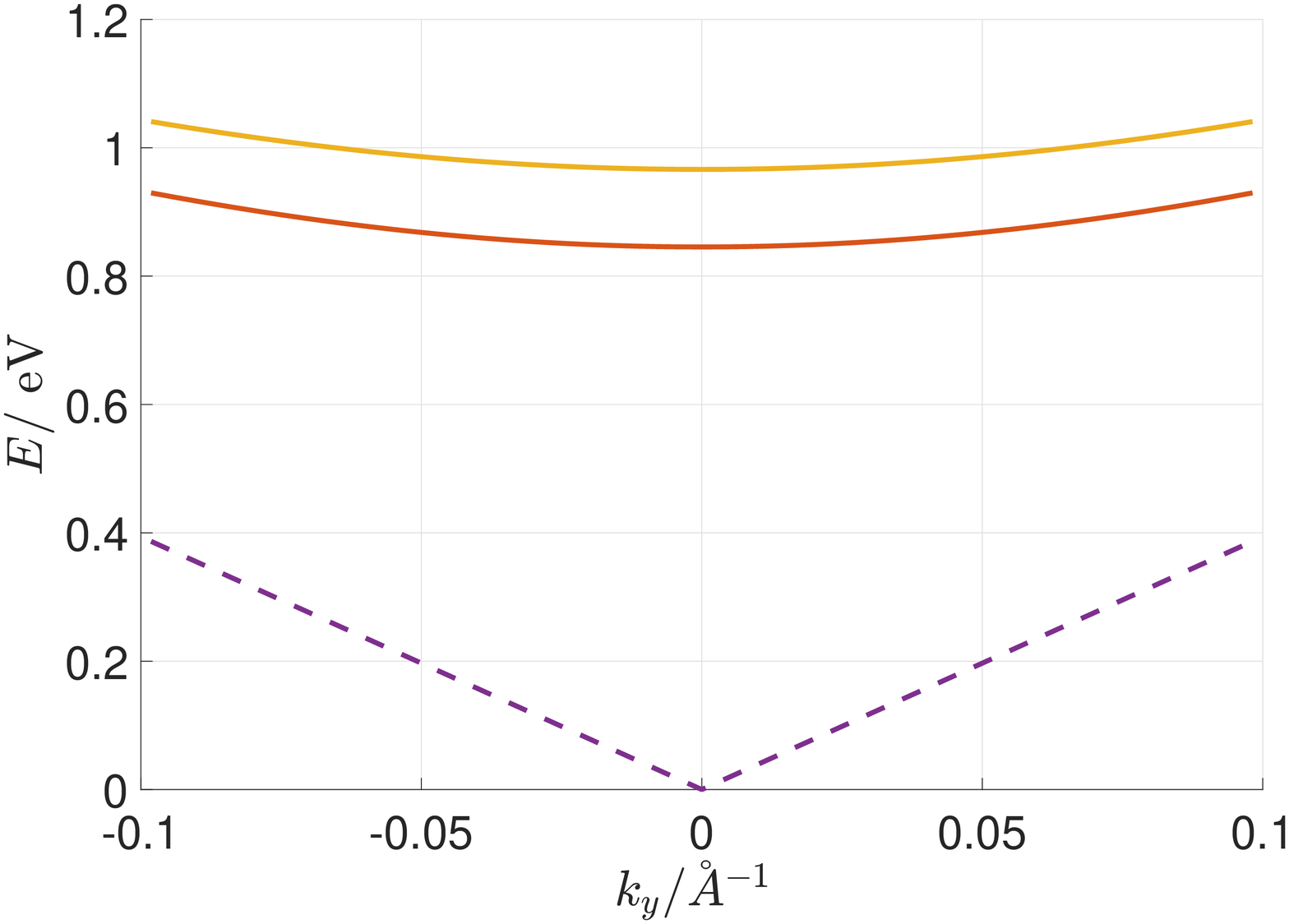}
        \label{fig:single particle E has ES}
    \end{subfigure}
    \hfill
    \begin{subfigure}[b]{0.475\textwidth}
        \caption[]{}%
        %{{\small Single valley subbands $\lambda \lambda_0=1$}}
        \centering
        \includegraphics[width=\textwidth]{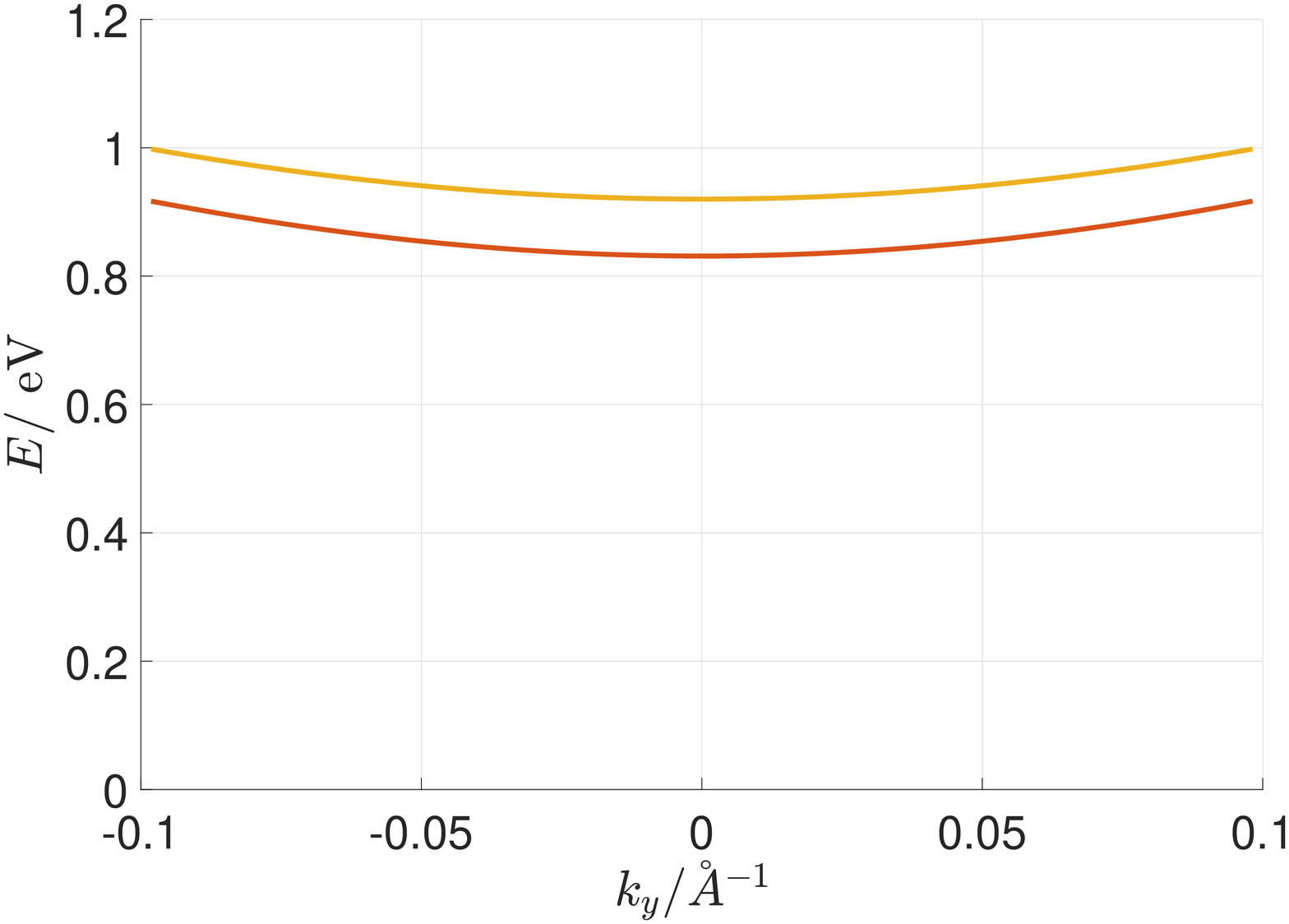}
        \label{fig:single particle E no ES}
    \end{subfigure}
    \caption[]
    {\small
    (a) Lowest three positive energy electric subbands for a single valley chiral fermion with two confined states (solid lines) and an edge state (dashed line). The vacuum and wire have opposite Chern number signs $\lambda \lambda_0=-1$.  The half gap $m = 0.8$eV, wire width is $L=50$ {\AA}, $v_F\hbar = 3.39$eV$\cdot$\AA.
    (b) Lowest two positive energy electric subbands for a single valley chiral fermion with no edge state. The vacuum and wire have the same Chern number signs, $\lambda \lambda_0=1$.  The half gap $m = 0.8$eV, wire width is $L=50$ {\AA}, $v_F\hbar = 3.39$eV$\cdot$\AA. }
    \label{fig:single particle E}
\end{figure*}

\begin{widetext}

The edge states correspond to evanescent solutions
of the Hamiltonian  equation (see Appendix A), with wavefunctions
\begin{equation}\label{eq: ribbon edge state}
  \vec{\psi}_{0, \vec{k}}^{\tau} (\vec{r}) = A_0(\tau; \vec{k})
    \left(
    \begin{array}{c}
        m+E \\
      i(\tau k_x + k_y) \\
    \end{array}
  \right)
  e^{-k_x x + ik_y y} + B_0(\tau; \vec{k})
    \left(
    \begin{array}{c}
        m+E \\
      i(-\tau k_x + k_y) \\
    \end{array}
  \right)
  e^{k_x x+ i k_y y}
\end{equation}
and
\begin{equation}\label{eq: A0}
  A_0(\tau) = \tau \sqrt{\frac{ ( (m + E)^2 - (\tau k_x-k_y)^2 ) (m+ k_x) } {8E(E+m)^2 L_y\left( m  - L (m^2- k_x^2) \right)}},
\end{equation}
\begin{equation}\label{eq: B0}
  B_0(\tau) = -\tau
        \sqrt{\frac{ ( (m + E)^2 - (\tau k_x+k_y)^2 ) (m- k_x) } {8E(E+m)^2 L_y\left( m  - L (m^2- k_x^2) \right)}},
\end{equation}
%{\bf is it necessary to define the energy of the evanescent modes?}
and energy
 \begin{equation}
  E = \sqrt{m^2- k_x^2+ k_y^2}.
\end{equation}
In these expressions  the evanescent wave vector $k_x$ satisfies the transcendental equation
\begin{equation}\label{eq: infinite mass evanescent kx}
  e^{-2 k_x L} = \frac{|m|- k_x}{|m| + k_x}.
\end{equation}
\end{widetext}
Note that Eq. \ref{eq: infinite mass evanescent kx} only has solutions when
the wire is wider than a minimal value ($L^*$), given by
\begin{equation}\label{eq: minimal width(main text)}
  L^* = \frac{\hbar v_F}{m}.
\end{equation}
where we have replaced the explicit functional dependence on $\hbar v_F$.  Examples of the single particle dispersions relevant to our model are given in Figs. \ref{fig:single particle E has ES} and \ref{fig:single particle E no ES}.

%%\FloatBarrier
%\begin{figure*}[!htb]
%  \centering
%  \begin{minipage}[b]{0.495\textwidth}
%    \includegraphics[width=\textwidth]{fig_draft1/SingleParticleEnergy/WSe2,5nm,has_ES.eps}
%    \caption{Lowest three positive energy electric subbands for a single valley chiral fermion with two confined states (solid lines) and an edge state(dashed line). The vacuum and ribbon have opposite Chern number signs $\lambda \lambda_0=-1$.  The half gap $m = 0.8$eV, ribbon width is $L=50$ {\AA}, $v_F\hbar = 3.393$eV.}
%    \label{fig:single particle E has ES}
%  \end{minipage}
%  \hfill
%  \begin{minipage}[b]{0.495\textwidth}
%    \includegraphics[width=\textwidth]{fig_draft1/SingleParticleEnergy/WSe2,5nm,no_ES.eps}
%    \caption{Lowest two positive energy electric subbands for a single valley chiral fermionwith no edge state. The vacuum and ribbon have the same Chern number signs, $\lambda \lambda_0=1$.  The half gap $m = 0.8$eV, ribbon width is $L=50$ {\AA}, $v_F\hbar = 3.393$eV.
%    }
%    \label{fig:single particle E no ES}
%  \end{minipage}
%\end{figure*}
%%\FloatBarrier

\section{Plasmon Wavefunctions, Energies, and Dipole Moments}
In our study we are interested in the intrinsic dipole moment of plasmon states of a one dimensional channel.  Most previous studies of these focus on the dielectric function, as computed in the random phase approximation (RPA) \cite{1989DasSarma,Das-Sarma:1985aa,Li:1991aa}.  This reveals the plasmon frequencies and their impact on the charge response of the system to applied electric fields.   For our purpose we need access to the plasmon wavefunctions. The approach we adopt casts the plasmon wavefunction as a linear combination of particle-hole excitations around a Fermi sea ground state. While equivalent to RPA, it is best understood as a time-dependent Hartree approximation.
In this section we explain how this approach is implemented and how the intrinsic transverse dipole moment may be extracted from it.

\subsection{Hamiltonian and Plasmon Raising Operator}
Our Hamiltonian is a sum of single particle and interaction parts, $\hat{H}_0+\hat{V}$.  The first of these is
\begin{equation}
\hat{H}_0=\sum_m\sum_{k_y}\sum_{\tau,s}E(m,k_y,\tau) c_{m,k_y,\tau,s}^{\dag} c_{m,k_y,\tau,s}
\label{H0}
\end{equation}
where $c_{m,k_y,\tau,s}^{\dag}$ creates an electron in subband $m$ with longitudinal momentum $k_y$, with valley and spin indices $\tau$ and $s$, respectively.  $E(m,k_y,\tau)$ is the single particle energy, as computed in the previous section.  For the interaction term we write
%\begin{widetext}
\begin{equation}
  \hat{V} = \frac{1}{2}\sum_{s,s'}\int d^2r\int d^2r' :\vec{\Psi}_{s}^{\dagger}(\vec{r}) \cdot
  \vec{\Psi}_{s}(\vec{r})
  V(\vec{r}- \vec{r}')
  \vec{\Psi}_{s'}^{\dagger}(\vec{r}') \cdot \vec{\Psi}_{s'}(\vec{r}') :
  \label{V}
\end{equation}
%  \end{widetext}
where the (vector) field operator $\vec{\Psi}_{s}^{\dagger}(\vec{r})$ creates an electron of spin $s$ in the two orbitals of the chiral fermion system at the location $\vec{r}$, and $:\hat{\mathcal{O}}:$ denotes normal ordering of an operator $\hat{\mathcal{O}}$.  Expanding in the single particle wire states,
\begin{equation}\label{eq: field operator}
  \vec{\Psi}_s(\vec{r})  = \sum_{n, k_y,\tau} c_{n, k_y,\tau,s}\vec{\psi}_{\vec{k},s}^{\tau} (\vec{r}),
\end{equation}
brings the interaction to a form which may be written as
\begin{widetext}
\begin{equation}
\hat{V} = \sum_{s,s'} \sum_{\substack{n_1,n_2,n_3,n_4 \\ k_{y1}, k_{y2}, k_{y3}, k_{y4}\\ \tau_1,\tau_2,\tau_3,\tau_4 }}
            V_{\vec{k}_1,\vec{k}_2,\vec{k}_3,\vec{k}_4}^{\tau_1,\tau_2,\tau_3,\tau_4}  c^{\dagger}_{\vec{k}_1,\tau_1,s'}c^{\dagger}_{\vec{k}_2,\tau_2,s}
            c^{}_{\vec{k}_3,\tau_3,s}c^{}_{\vec{k}_4,\tau_4,s'},
\end{equation}
where $V_{\vec{k}_1,\vec{k}_2,\vec{k}_3,\vec{k}_4}^{\tau_1,\tau_2,\tau_3,\tau_4} = {1 \over 2}\int d^2r\int d^2r'
\vec{\psi}_{\vec{k}_1}^{\tau_1} (\vec{r})^*\cdot \vec{\psi}_{\vec{k}_4}^{\tau_4} (\vec{r}) V(\vec{r}-\vec{r}\,') \vec{\psi}_{\vec{k}_2}^{\tau_2} (\vec{r}\,')^*\cdot \vec{\psi}_{\vec{k}_3}^{\tau_3}(\vec{r}\,')$.
In writing this we have taken note of the fact that for each subband there is a single quantized transverse momentum magnitude, $k_x(n)$, whose states with positive and negative values are admixed to form the transverse states discussed in the last section.  Identifying $\vec{k} \equiv (k_x(n),k_y)$, this allows us to adopt a simplified indexing for the annihilation operators, $c_{\vec{k},\tau,s} \equiv c_{n,k_y,\tau,s}$.

In what follows we adopt a contact interaction $V(\vec{r} - \vec{r}') = u_0 \delta(\vec{r} - \vec{r}')$.  This yields intrasubband plasmon modes that disperse linearly with longitudinal plasmon momentum $K_y$.  If a $1/r$ potential is instead used, one expects to find $\omega(K_y) \sim K_y \ln K_y$ for at least one gapless plasmon mode; however in practice the divergence of the slope is extremely difficult to see \cite{1989DasSarma}.  Thus the contact interaction introduces significant simplification in the computation of the matrix elements, without loss of any essential qualitative behavior in the plasmon mode.  In practice, we choose the value of $u_0$ to match results for the slope of a plasmon mode as computed using the Coulomb interactions in a graphene system \cite{2017Karimi}.
%Further detail on this matching procedure is provided in Appendix XXX.

Collective excitation of the system  can be obtained from operators satisfying the equation \cite{THOULESS:1961aa,Sawada:1957aa}
\begin{equation}\label{eq: plasmon frequency, bv}
  [\hat{H}, \hat{Q}^{\dagger}_{K_y}] =\hbar  \omega_{K_y} \hat{Q}^{\dagger}_{K_y}.
\end{equation}
In general analytic solutions to Eq. \ref{eq: plasmon frequency, bv} are not available. However in the case of plasmons, corresponding to charge density excitations in the system,  one may approximate the form of the plasmons raising operator $\hat{Q}^{\dagger}_{K_y}$ as a linear combination of single particle-hole pairs \cite{Sawada:1957aa},
\begin{equation}\label{def: plasmon creation, bv} %bv stands for both valleys
  \hat{Q}^{\dagger}_{K_y} \equiv \sum_{m_1,m_2, k_y',\tau,s} a_{m_1,m_2,k_y',\tau}(K_y) c^{\dagger}_{m_1,K_y+k_y',\tau,s} c_{m_2,k_y',\tau,s},
\end{equation}
and then treat the commutator $[\hat{V},\hat{Q}^{\dagger}_{K_y}]$ in the time-dependent Hartree approximation,
\begin{align}\label{eq: potential commutator RPA,bv}
[\hat{V}, \hat{Q}^{\dagger}_{K_y}]  \approx & 2 \sum_{s,s'}\sum_{n_1,n_2,n_3,n_4}\sum_{\tau_1,\tau_4,\tau} \sum_{k_{y1},k_y'} V_{n_1,n_2,n_3,n_4}^{\tau_1,\tau,\tau,\tau_4}(K_y+k_y', k_{y1},K_y+k_{y1}, k_y') a_{n_3,n_2,k_{y1},\tau,s}(K_y) \nonumber\\
& \times c^{\dagger}_{n_1,K_y+k_y',\tau_1,s'} c_{n_4,k_y',\tau_4,s'} (n_{F}(n_2,k_{y1},\tau) - n_{F}(n_3,K_y+k_{y1},\tau) ).
\end{align}
Together with the commutator involving the single particle Hamiltonian $\hat{H}_0$, one arrives at an eigenvalue equation for the particle-hole weights $a_{n_1,n_2,k_y,\tau,s}(K_y)$ and the plasmon frequency $\omega(K_y)$,
\begin{align}\label{eq: ribbon RPA eigen eq, bv}
&2 \sum_{n_1,n_2} \sum_{k_{y1}}\sum_{\tau,s}  V_{n_1',n_2,n_1,n_2'}^{\tau', \; \tau, \;\; \tau, \;\; \tau'}(K_y+k_y', k_{y1},K_y+k_{y1}, k_y')
                                a_{n_1,n_2,k_{y1},\tau,s}(K_y) (n_{F}(n_2,k_{y1},\tau) - n_{F}(n_1,K_y+k_{y1},\tau) ) \nonumber\\
&= \left[\omega_{K_y} - E(n_1',K_y+k_y',\tau') + E(n_2',k_y',\tau') \right] a_{n_1',n_2',k_y',\tau',s'}(K_y).
\end{align}
{In this work we work strictly in the zero temperature limit, so that $n_F(n,k_{y},\tau) = 1(0)$ if the state $(n,k_y,\tau)$ is occupied (unoccupied) in the ground state.}

We solve Eq. \ref{eq: ribbon RPA eigen eq, bv} numerically by retaining a discrete set of points in the $k_y$ sum, so that it becomes a matrix eigenvalue equation.  Because we are interested in the lowest lying plasmon modes, we further simplify the equation by retaining only intra-band particle-hole excitations, so that we take $a_{n_1,n_2,k_y,\tau,s}(K_y) \ne 0$ only when $n_1=n_2$; we have verified that keeping inter-subband excitations has little effect on our results.
{We have further verified that increasing the number of $k_y$ points used for the results reported below have little effect on them.}
%\end{widetext}

%By discretizing the $y$ direction momentum on a grid we numerically solve the above eigen equation for intraband plasmon. Due to the vastly different energy scale for inter and intra band plasmons, we may consider them separately, but for this paper we consider only the intraband plasmon.

%In order to perform the numerical computation, we determine the strength of the contact interaction by fitting the slope of the highest single subband plasmon mode near $K_y=0$ in graphene nanoribbon to the numerical computation performed using Coulomb interaction in 41-aGNR\cite{2017Karimi}, we find the effective interacting strength for the contact potential $V(\vec{r}-\vec{r}') = u_0 \delta(\vec{r}-\vec{r}')$ to be $u_0 = 4.1537\times 10^{2}$ eV$\cdot$ \AA$^2$ (For more details, see supplementary).

\subsection{Plasmon Dipole Moment}
In previous work \cite{2021excitonQGD,2021PlasmonQGD} we demonstrated that two-body excitations, including excitons and plasmons, may carry an internal dipole moment that is tied to the quantum geometry of their wavefunctions.
One sees this by defining Berry connections specific for the electrons
($\alpha$=1) and holes ($\alpha$=2),
\begin{equation*}
{\mathbfcal A } ^{(\alpha)} (\Kn ) = i \langle u_{\Kn,\alpha} | \vec {\nabla} _{\Kn} |u_{\Kn,\alpha}\rangle \end{equation*}
with
\begin{equation*}
|u_{\Kn,\alpha}\rangle = e ^{i {\Kn} {\rn}_\alpha} |\Phi _{\Kn}\rangle \, \, ,
\end{equation*}
where $|\Phi _{\Kn}\rangle$ is the wavefunction of the excited state.  These connections
can be directly related to the average electric dipole moment ${\bf d}$ of a plasmon,
%\begin{widetext}
\begin{eqnarray}
{\bf d}& = &  e <\Phi_{\Kn} |\rn _1 - \rn _2 | \Phi _{\Kn}>  \nonumber \\
%& = &
% i e \left[\langle
%\Phi_{\bf K} | e^{i{\bf K}\cdot{\bf r_1}} \left( \vec{\nabla}_K e^{-i{\bf K}\cdot{\bf r_1}} \right) |\Phi_{\bf K}\rangle
%- \langle \Phi_{\bf K} | e^{i{\bf K}\cdot{\bf r_2}}
%\left( \vec{\nabla}_K e^{-i{\bf K}\cdot{\bf r_2}} \right) | \Phi_{\bf K} \rangle
%\right] \nonumber \\
&= &  i e \left[\langle u_{\bf K ,1} |\vec{\nabla}_K | u_{\bf K,1 }\rangle - \langle u_{\bf K,2} |\vec{\nabla}_K | u_{\bf K,2 } \rangle \right]
% =i e \sum _{qn} U^*(\Kn,\qn) \nabla _{\qn} U(\Kn,\qn)
=e\left[{\mathbfcal A}^{(1)}({\bf K})- {\mathbfcal A}^{(2)}({\bf K}) \right] \, \equiv e {\mathbfcal D}(\Kn),
\end{eqnarray}
where ${\mathbfcal D}$ is the quantum geometric dipole.
This quantity is relevant to plasmons because they may be understood as particle-hole excitations around a Fermi surface.  In a two-dimensional system one finds ${\mathbfcal D}(\Kn)$ is orthogonal to $\Kn$, and for small $K$ it is linear in $K$.  This geometry suggests that when plasmons carry a non-vanishing ${\mathbfcal D}$ in a two-dimensional material, plasmons confined to a one-dimensional channel of the same system may carry a transverse dipole moment.  We can check this by computing the plasmon dipole moment directly.
For a wire oriented along the $\hat{y}$-direction, following the reasoning above, for a plasmon state $|\Phi_{K_y}\rangle$ with momentum $K_y$ along the wire one may write
\begin{equation}\label{eq: dipole as average pos}
  \mathcal{D}_x(K_y) = \mathcal{A}^{(1)}_x- \mathcal{A}^{(2)}_x = \langle \Phi_{K_y} |x_e - x_h| \Phi_{K_y} \rangle.
\end{equation}
Recalling the notation above in which a vector $\vec{k}=(k_x(n),k_y)$ specifies an electron state with longitudinal momentum $k_y$ in a transverse state $n$, we write $\vec{\psi}_{\vec{k}}^{\tau} \rightarrow \vec\psi_{n,k_y}^{\tau}$, yielding an explicit expression,
%\begin{widetext}
\begin{align}\label{eq: dipole explicit}
  \mathcal{D}_x(K_y) &= \sum_{\substack{m_1,m_2,\\m_1',m_2'}}\sum_{ k_y,\tau,s} a^*_{m_1',m_2',k_y,\tau,s}(K_y)a_{m_1,m_2,k_y,\tau,s}(K_y)  \nonumber\\
  &\times \left( \delta_{m_2,m_2'}\int x \vec\psi^{\tau *}_{m_1',k_y+K_y}(\vec{r}) \cdot \vec\psi^{\tau}_{m_1,k_y+K_y}(\vec{r})  d^2r
        -  \delta_{m_1,m_1'}\int x \vec\psi^{\tau}_{m_2',k_y} (\vec{r})  \cdot \vec\psi^{\tau *}_{m_2,k_y}(\vec{r})  d^2r   \right).
\end{align}
%\end{widetext}
In our numerical calculations, Eq. \ref{eq: dipole explicit} is used to compute the dipole moment of a plasmon state.  As we shall see, one finds that plasmons of a single chiral Dirac fermion {nanowire} {\it generically} have non-vanishing $\mathcal{D}_x(K_y)$, but with increasing wire width, this vanishes unless the corresponding two-dimensional system has a non-vanishing quantum geometric dipole.

\end{widetext}

\section{Results}
\FloatBarrier
In this paper we focus on intraband plasmons, which for our contact interaction disperse linearly with momentum from zero energy.  In general we find that the number of such gapless plasmon modes is equal to the number of subbands which cross the Fermi energy, all of which may carry non-vanishing transverse dipole moments.  We begin by considering the computationally simplest case of a single chiral fermion flavor.  In principle such a system might be created on the surface of a topological insulator infused with ferromagnetically ordered dopants that gap the surface states everywhere except in a narrow channel, where plasmons can be hosted.  We discuss more common cases involving nanoribbons of van der Waals materials further below, for which the effects of multiple valleys and time-reversal symmetry have important consequences.

\subsection{Single Chiral Fermion}
Our model Hamiltonian for a single chiral fermion is $\hat{H}=\hat{H}_0+\hat{V}$, with $\hat{H}_0$ and $\hat{V}$ given by Eqs. \ref{H0} and \ref{V}, respectively, in which only a single valley flavor $\tau$ is retained.  For such systems we need to choose whether the vacuum outside the same system has the same or opposite Chern number as the one-dimensional system, i.e. whether $\lambda\lambda_0 = 1$ or $-1$, as discussed in Section II.  This determines whether the wire hosts edge states.  For the realization described above one may toggle between the two cases by flipping the direction of the magnetic impurities defining the channel.  The qualitative behavior of the system turns out to be the same irrespective of whether the wire hosts edge states.

%
%\begin{figure}[!htb]
%  \centering
%  \begin{minipage}[b]{0.495\textwidth}
%    \includegraphics[width=\textwidth]{fig_draft1/Dipole_degeneracy/Energy_ES_off_4+0_K_valley_EF=0.80556eV_gap=0.4eV,L=60A.eps}
%    \caption{Intraband plasmon energy for K valley excitation where 4 subbands are kept(the edge state is turned off due to boundary condition). There are 4 plasmon modes in total. The parameters are $L =   60$ {\AA},$\Delta =    0.4$ eV,$v_F\hbar =    3.939$ eV$\cdot${\AA},Fermi energy $E_F = 1.00695$ eV.Gap is one quarter of WSe$_2$. }
%    \label{fig: Dipole break degeneracy ES off, Energy}
%  \end{minipage}
%  \hfill
%  \begin{minipage}[b]{0.495\textwidth}
%    \includegraphics[width=\textwidth]{fig_draft1/Dipole_degeneracy/Dipole_ES_off_4+0_K_valley_EF=0.80556eV_gap=0.4eV,L=60A.eps}
%    \caption{Intraband plasmon dipole for K valley excitation where 4 subbands are kept.
%    The parameters are $L =   60$ {\AA},$\Delta =    0.4$ eV,$v_F\hbar =    3.939$ eV$\cdot${\AA},Fermi energy $E_F = 1.00695$ eV. Gap is quarter of WSe$_2$.}
%    \label{fig: Dipole break degeneracy ES off, Dipole}
%  \end{minipage}
%\end{figure}

%\textbf{(I think we want to retain two seperate figures for this first result.  After this we would like to somehow combine the plasmon dispersions and the accompanying dipoles into single figures.)}

We begin with typical results, illustrated in Figs. \ref{fig: Dipole break degeneracy ES off, Energy} and \ref{fig: Dipole break degeneracy ES off, Dipole} for boundary conditions in which there are no edge states ($\lambda\lambda_0=1$.) One observes several gapless plasmons, which at long wavelengths disperse linearly with momentum, as expected for this model.  The gapless modes illustrated all lie above the energies of the particle-hole continuum.  In general, the number of gapless modes is equal to the number of occupied subbands;  we demonstrate this explicitly for the case of two occupied subbands in Appendix B. Importantly, all the plasmon modes exhibit non-vanishing transverse dipole moments, with magnitudes proportional to the plasmon momentum.  As we discuss below, while this behavior is consistent with the (two-dimensional) quantum geometric properties of the system hosting the wire, it can be present in the wire geometry even when absent in the corresponding two-dimensional system.  We also find that such qualitative results are unaffected by edge states ($\lambda\lambda_0<0$), as illustrated in Figs. \ref{fig: Dipole break degeneracy ES on, Energy} and \ref{fig: Dipole break degeneracy ES on, Dipole}; the transverse dipole moment does not appear to depend on this aspect of the system topology.

%\begin{figure}[!htb]
%  \centering
%  \begin{minipage}[b]{0.495\textwidth}
%    \includegraphics[width=\textwidth]{fig_draft1/Dipole_degeneracy/Energy_ES_on_4+1_K_valley_EF=1.12778eV_gap=0.4eV,L=60A.eps}
%    \caption{Intraband plasmon energy for K valley excitation where 5(=4+1) subbands(including the edge state) are kept. There are 5 plasmon modes in total. The parameters are $L =   60$ {\AA},$\Delta =    0.4$ eV,$v_F\hbar =    3.939$ eV$\cdot${\AA},Fermi energy $E_F = 1.00695$ eV.Gap is one quarter of WSe$_2$. }
%    \label{fig: Dipole break degeneracy ES on, Energy}
%  \end{minipage}
%  \hfill
%  \begin{minipage}[b]{0.495\textwidth}
%    \includegraphics[width=\textwidth]{fig_draft1/Dipole_degeneracy/Dipole_ES_on_4+1_K_valley_EF=1.12778eV_gap=0.4eV,L=60A.eps}
%    \caption{Intraband plasmon dipole for K valley excitation where 5 subbands( including the edge state) are kept.
%    The parameters are $L =   60$ {\AA},$\Delta =    0.4$ eV,$v_F\hbar =    3.939$ eV$\cdot${\AA},Fermi energy $E_F = 1.00695$ eV. Gap is quarter of WSe$_2$.}
%    \label{fig: Dipole break degeneracy ES on, Dipole}
%  \end{minipage}
%\end{figure}

\begin{figure*}[!htb]
	\centering
    \begin{adjustbox}{minipage=\linewidth,scale=1.0}
	\begin{subfigure}[b]{0.495\textwidth}
		\caption[]%
		{{}}
		\centering
		\includegraphics[width=\textwidth]{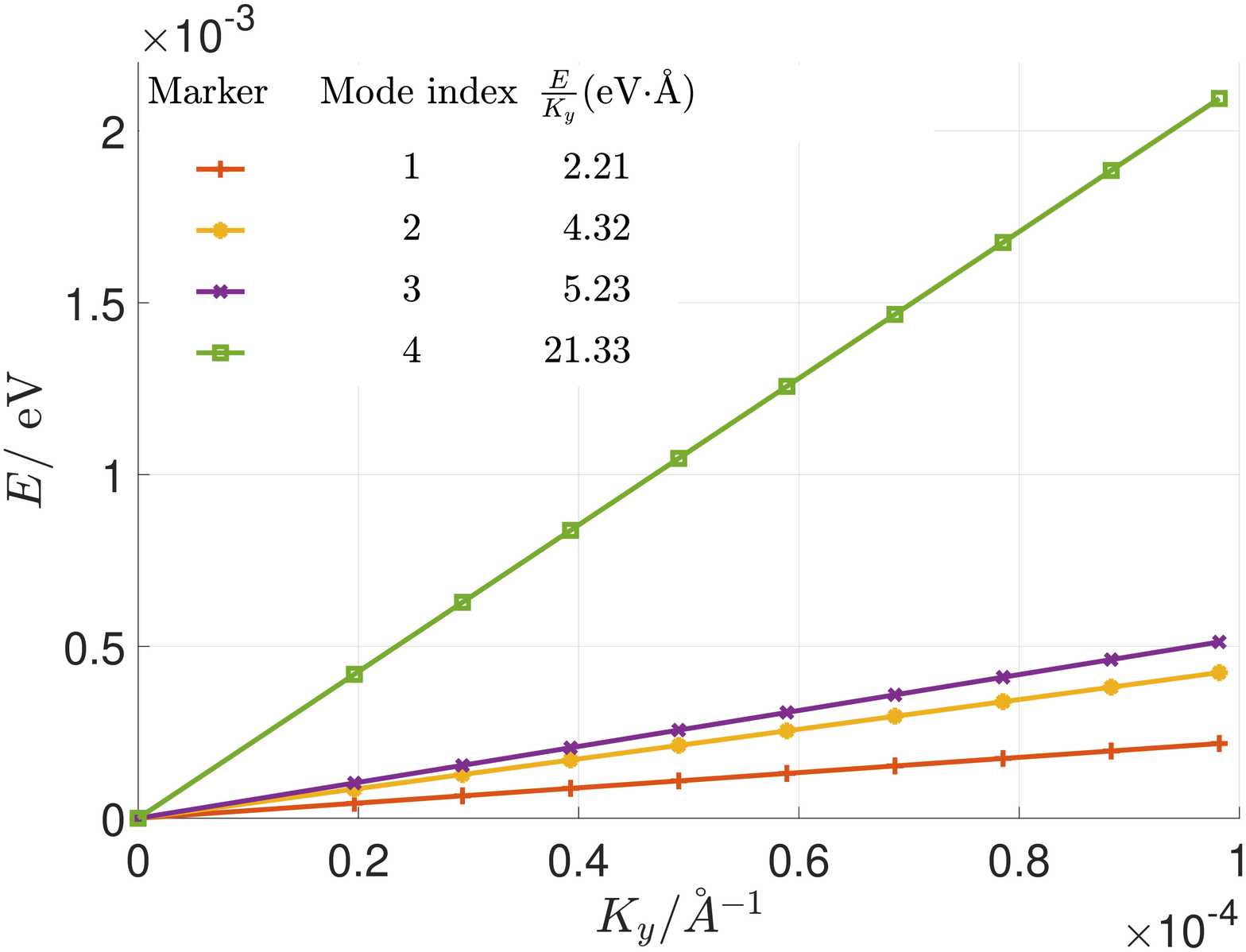}
		\label{fig: Dipole break degeneracy ES off, Energy}
	\end{subfigure}
	\hfill
	\begin{subfigure}[b]{0.495\textwidth}
		\caption[]%
		{{}}
		\centering
		\includegraphics[width=\textwidth]{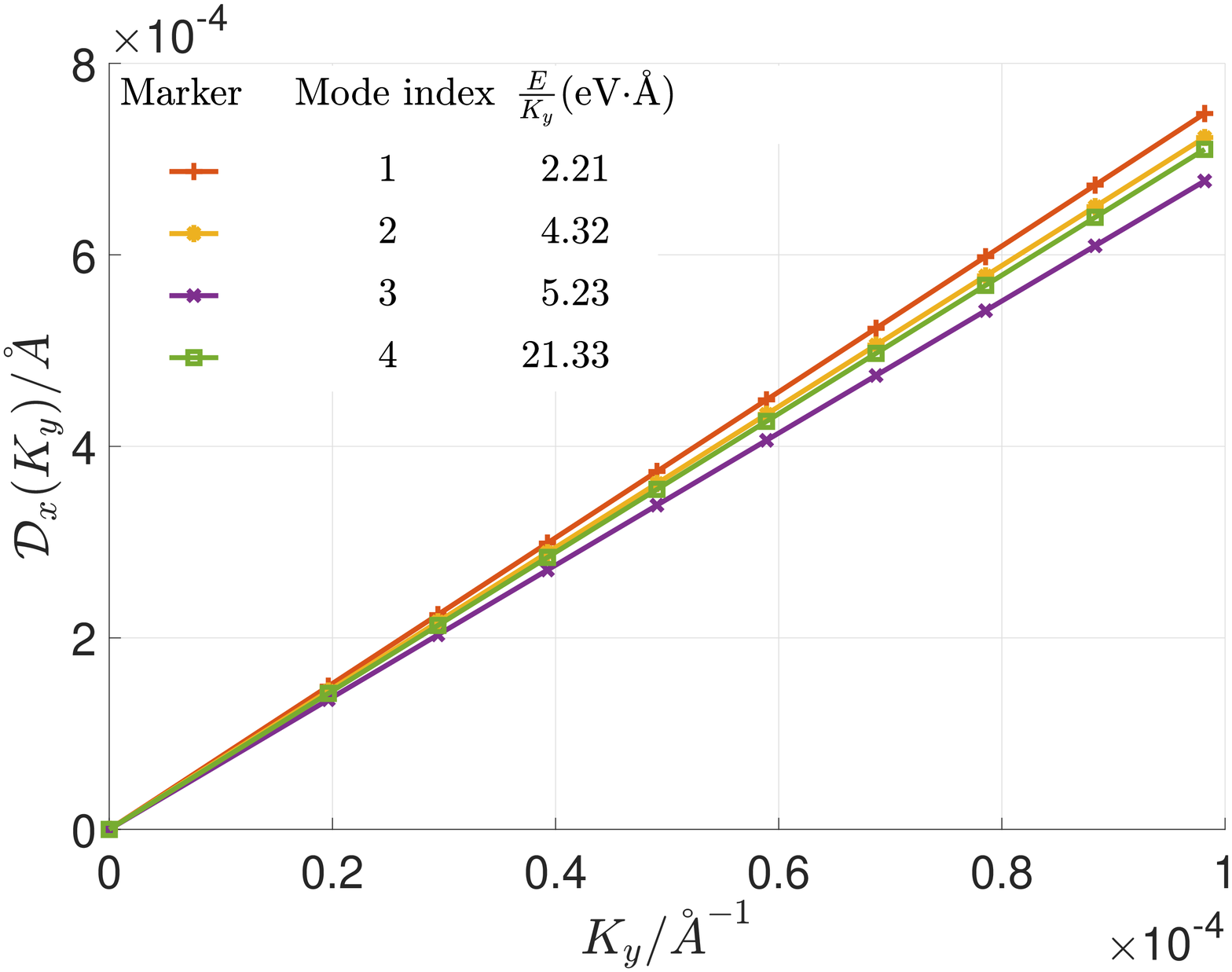}
		\label{fig: Dipole break degeneracy ES off, Dipole}
	\end{subfigure}
	\vskip\baselineskip
	\begin{subfigure}[b]{0.495\textwidth}
		\caption[]%
		{{}}
		\centering
		\includegraphics[width=\textwidth]{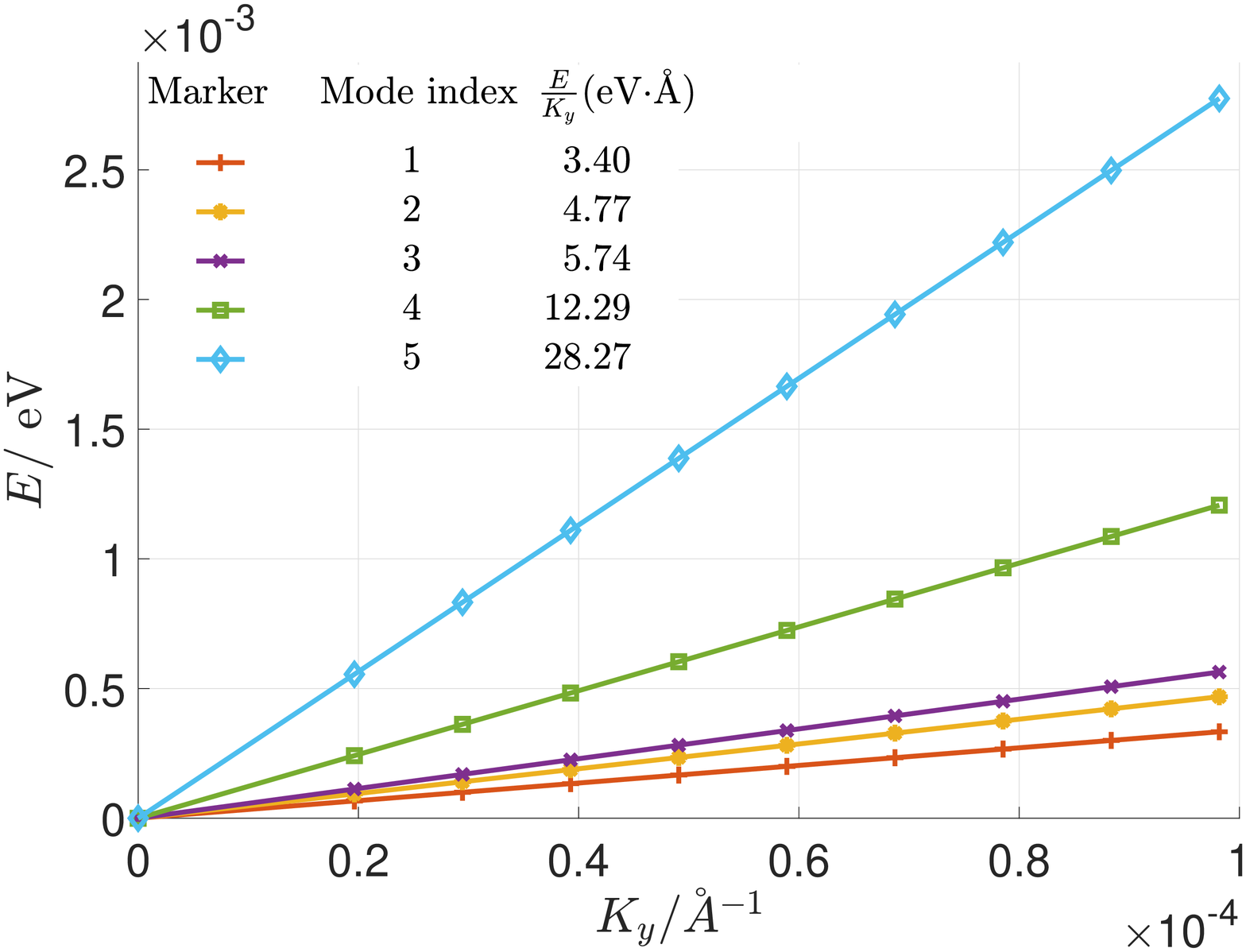}
		\label{fig: Dipole break degeneracy ES on, Energy}
	\end{subfigure}
	\hfill
	\begin{subfigure}[b]{0.495\textwidth}
		\caption[]%
		{{}}
		\centering
		\includegraphics[width=\textwidth]{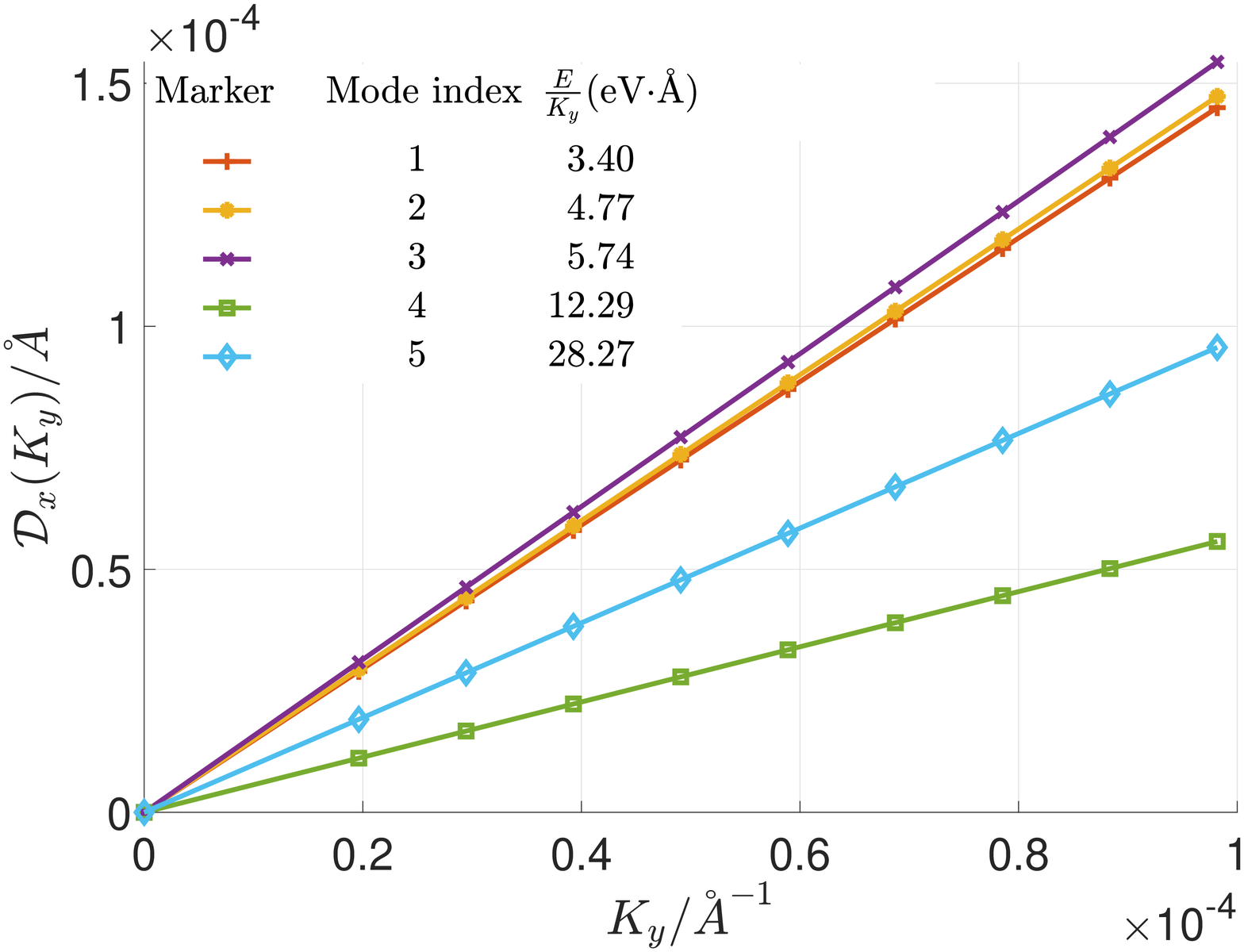}
		\label{fig: Dipole break degeneracy ES on, Dipole}
	\end{subfigure}
    \end{adjustbox}
	\caption[]
	{Intraband plasmon energies and transverse dipole moments for $K$ valley excitations. The wire width is $L =   60$ {\AA}, assumed gap $\Delta =    0.4$ eV, $v_F\hbar =    3.94$ eV$\cdot${\AA}, Fermi energy $\varepsilon_F = 1.007$ eV for (a) and (b),
    Fermi energy $\varepsilon_F = 1.128$ eV for (c) and (d), .
    (a) Plasmon energies when 4 subbands are occupied (no edge state due to boundary condition). There are 4 plasmon modes in total.
    (b) Transverse dipole moments corresponding to (a).
    (c) Plasmon energies when 5(=4+1) subbands (including an edge state) are occupied.
    (d) Transverse dipole moments corresponding to (b).
    }
	\label{fig:energy dipole curve for m=0.4 eV}
\end{figure*}

\begin{figure*}[!htb]
	\centering
    \begin{adjustbox}{minipage=\linewidth,scale=1.0}
	\begin{subfigure}[b]{0.495\textwidth}
		\caption[]%
		{{}}
		\centering
		\includegraphics[width=\textwidth]{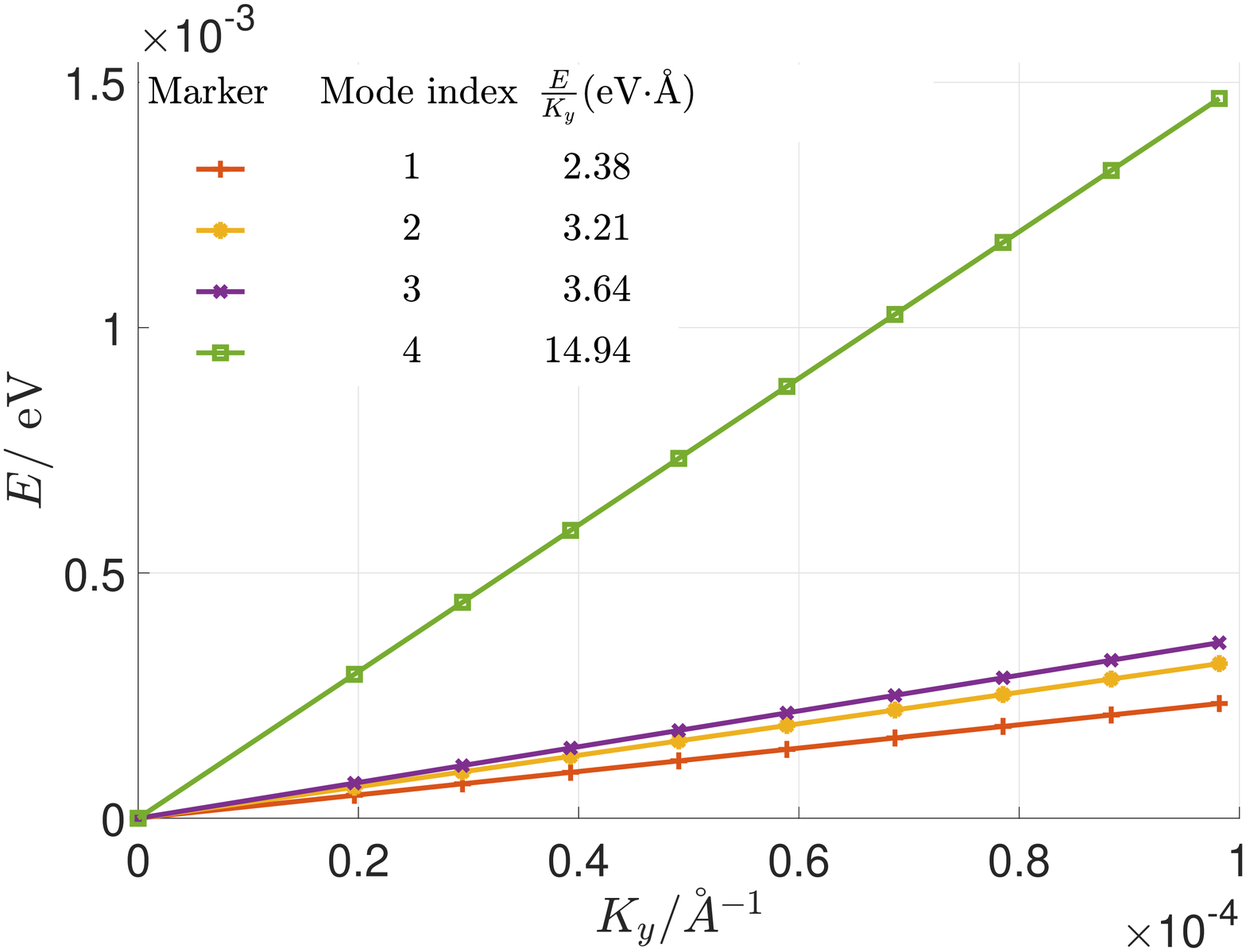}
		\label{fig:Intra energy K valley ES off 4 bands}
	\end{subfigure}
	\hfill
	\begin{subfigure}[b]{0.495\textwidth}
		\caption[]%
		{{}}
		\centering
		\includegraphics[width=\textwidth]{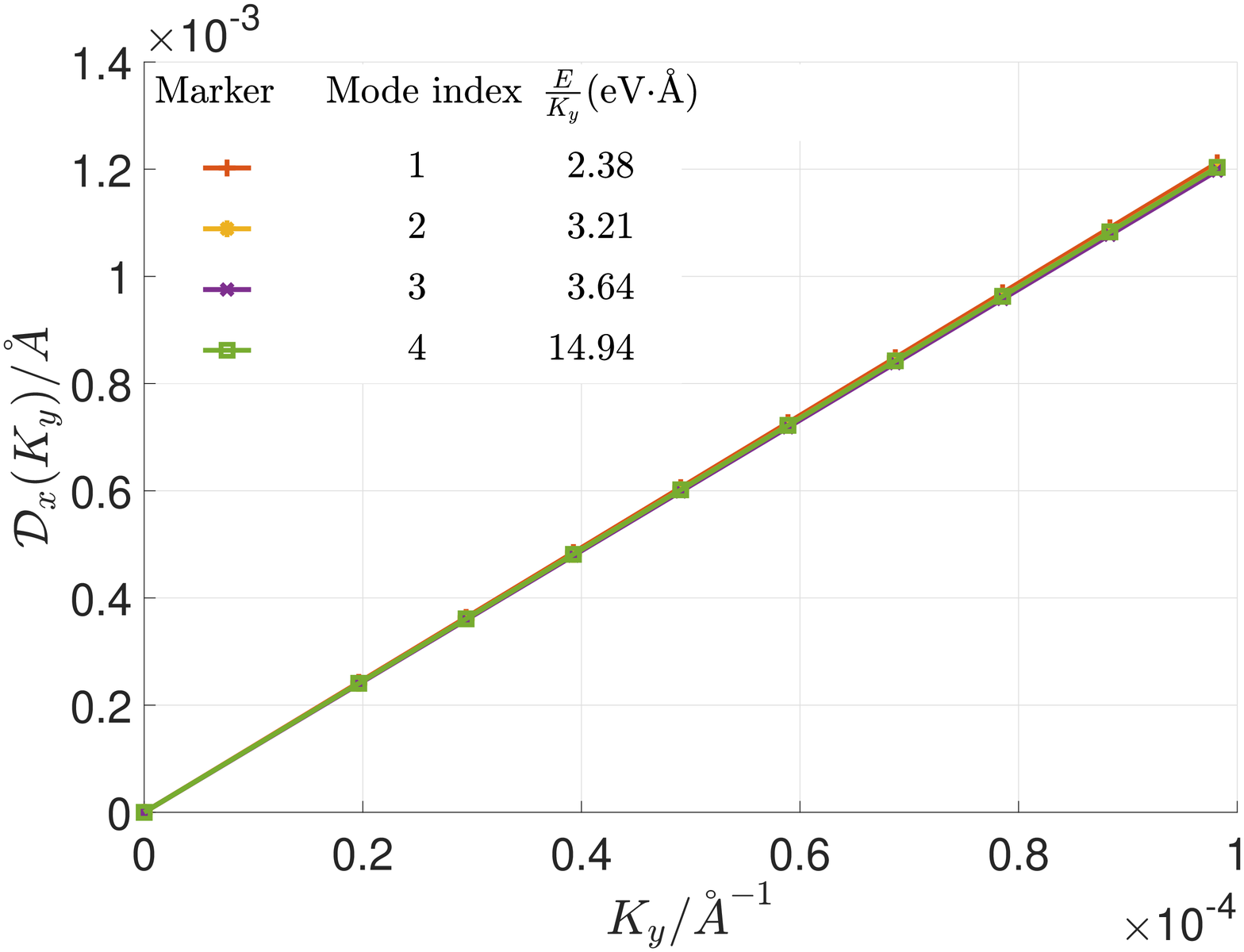}
		\label{fig:Intra dipole K valley ES off 4 bands}
	\end{subfigure}
	\vskip\baselineskip
	\begin{subfigure}[b]{0.495\textwidth}
		\caption[]%
		{{}}
		\centering
		\includegraphics[width=\textwidth]{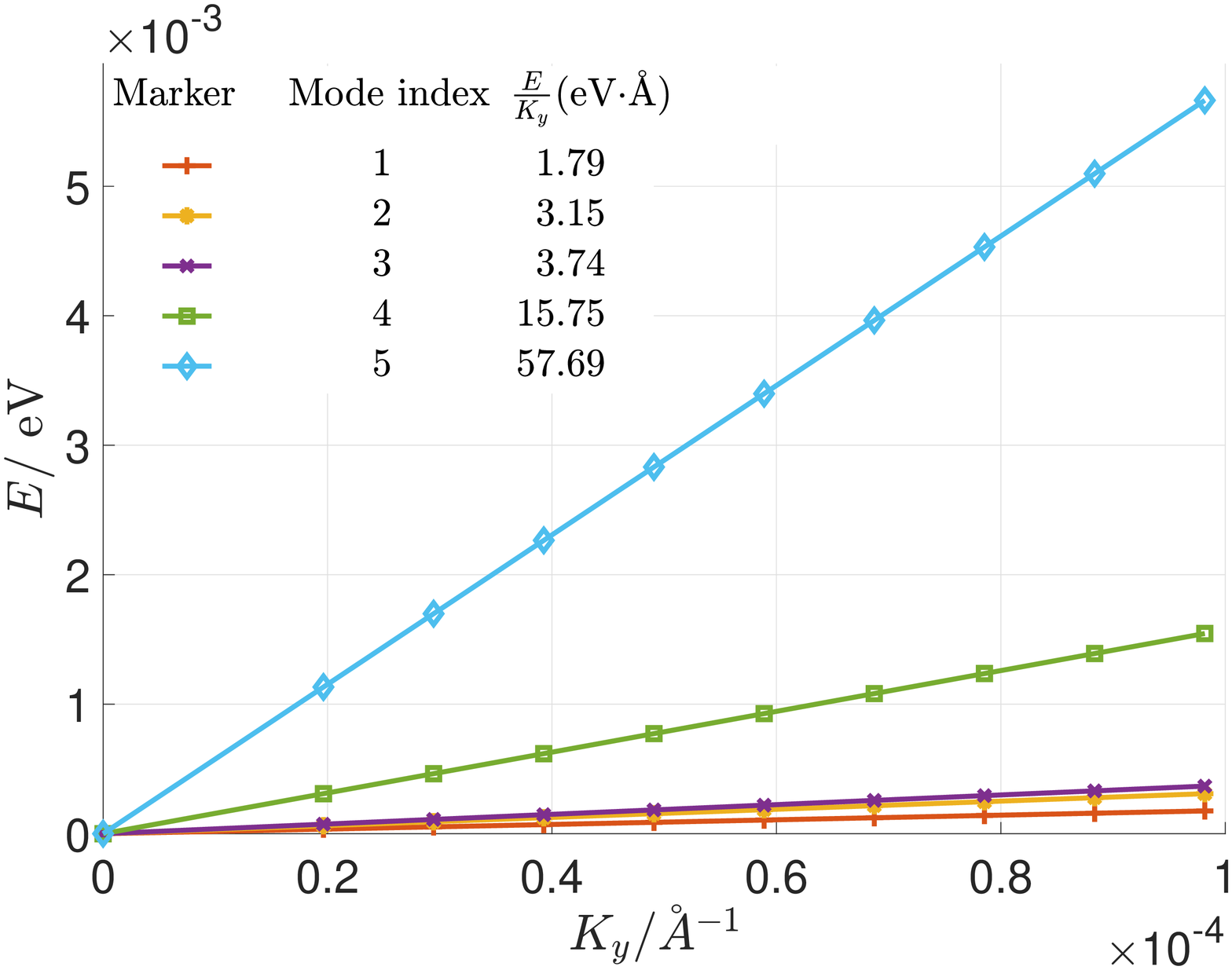}
		\label{fig:Intra energy K valley ES on 4 bands}
	\end{subfigure}
	\hfill
	\begin{subfigure}[b]{0.495\textwidth}
		\caption[]%
		{{}}
		\centering
		\includegraphics[width=\textwidth]{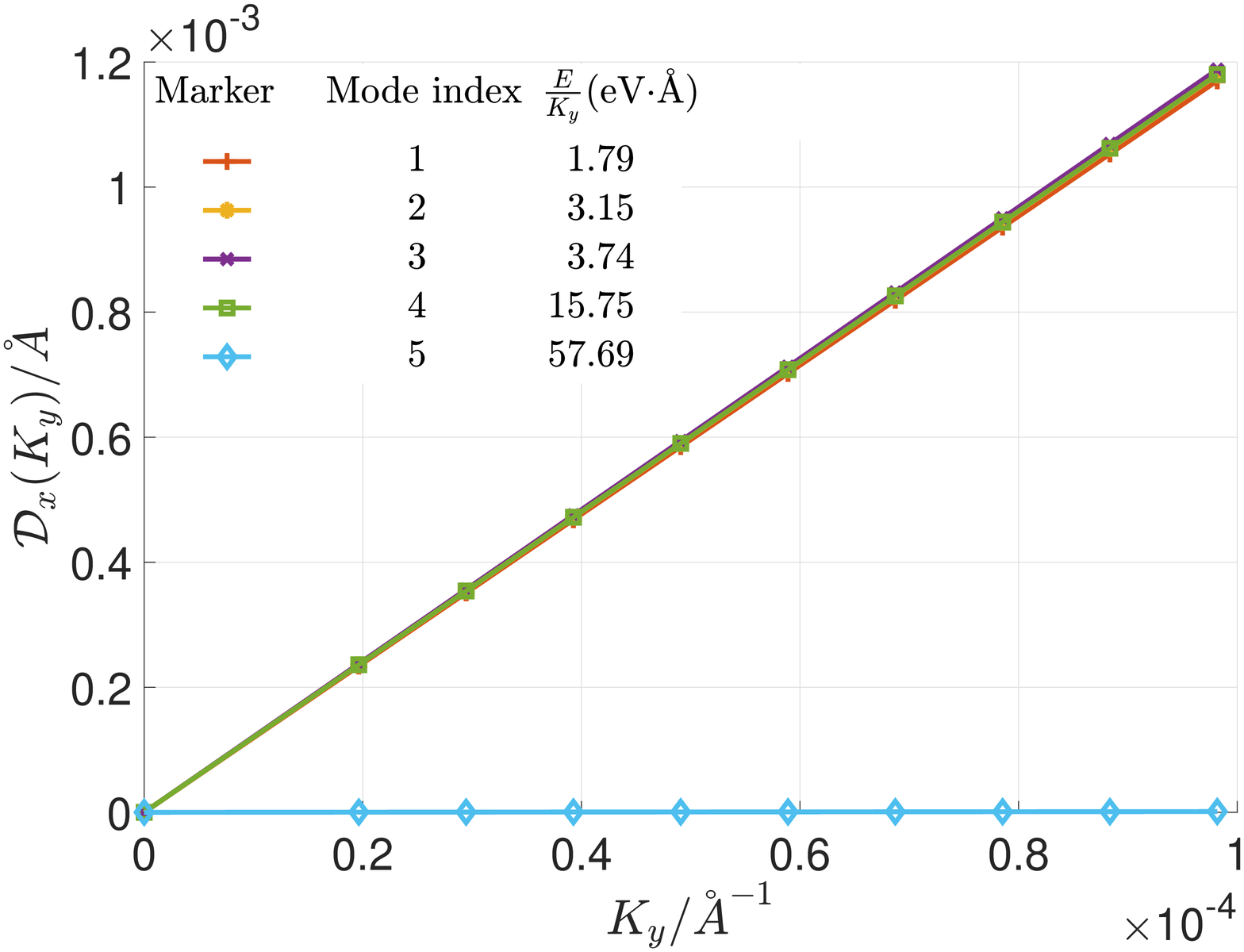}
		\label{fig:Intra dipole K valley ES on 4 bands}
	\end{subfigure}
    \end{adjustbox}
	\caption[]
	{\small
    Intraband plasmon energy and transverse dipole moments for $K$ valley excitations. Parameters are given by $L =   86$ {\AA}, $\Delta =    1.6$ eV, $v_F\hbar =    3.94$ eV$\cdot${\AA}, Fermi energy $\varepsilon_F = 1.007$ eV.
    (a) Intraband plasmon energies when 4 subbands are occupied.
    (b) Transverse dipole moments corresponding to (a).
    (c) Intraband plasmon energies when 5 modes are occupied (including edge states).
    (d) Transverse dipole moments corresponding to (c).
    }
	\label{fig:Intra energy and dipole for K valley 4 bands}
\end{figure*}

Surprising behavior of the transverse dipole moment emerges for systems with relatively large gaps.
Fig. \ref{fig:Intra energy K valley ES off 4 bands} illustrates such a case, in which the Hamiltonian parameters have been chosen to model a single valley of WSe$_2$ \cite{Xiao_2012}, and $\lambda\lambda_0=1$ (no edge states on the wire).
Fig. \ref{fig:Intra dipole K valley ES off 4 bands} illustrates the transverse dipole moment of these plasmon modes. Remarkably, one finds essentially the same value for all the modes.  This can be understood by close examination of the expression for the plasmon dipole for small momentum $K_y$, which we show in Appendix C to have the form
\begin{widetext}
\begin{align}\label{eq: dipole explicit intraband, explicit}
  \mathcal{D}_x(K_y)
& = \sum_{\tau}\frac{\tau K_y  (L m + \lambda_0 \tau ) }{2 \varepsilon_F^3}
\sum_{n}\sum_{ k_y,s}
\frac{ |a_{n,n,k_y,\tau,s}(K_y)|^2  }{  L +  \frac{\lambda_0 \tau m}{m^2+ k_x(n)^2}  } + \mathcal{O}(K_y^2).
\end{align}
\end{widetext}
where $k_x(n)$ is the quantized transverse momentum of the $n^{\rm th}$ subband.  Note that in this expression, in the present case where we consider a single valley (indexed by $\tau$), one may set $ \lambda\lambda_0 = \tau\lambda_0 $.
In situations where the gap parameter $m$ is large, the last term in the denominator becomes negligible,
so that the remaining sum $\sum_{n}\sum_{ k_y,\tau,s}
|a_{n,n,k_y,\tau,s}(K_y)|^2$ is determined {\it only} by the normalization of the plasmon wavefunction, and the resulting dipole moment becomes independent of the specific plasmon mode.

Figs. \ref{fig:Intra energy K valley ES on 4 bands} and \ref{fig:Intra dipole K valley ES on 4 bands} illustrate the corresponding results for the same parameters, but with $\lambda\lambda_0=-1$.  In this case the systems hosts edge states in addition to the confined single-particle states, so that there are five occupied subbands.  Here all but one of the plasmon modes host the same non-vanishing dipole moment, while the remaining mode does not. The result again can be understood from Eq. \ref{eq: dipole explicit intraband, explicit}.  In this case one finds that the mode with vanishing dipole moment has nearly all its weight in the edge state, for which $k_x(n) \approx im$, so that the denominator becomes divergent.  More physically, because of the relatively large gap, the penetration length of the edge state into the bulk becomes independent of $k_y$, as does the single-particle transverse wavefunction.  In this case the plasmon cannot sustain a transverse dipole moment. It is interesting to note that
the difference in behaviors apparent in Figs. \ref{fig:Intra dipole K valley ES off 4 bands} and
\ref{fig:Intra dipole K valley ES on 4 bands} in principle offers an interesting way to distinguish when the one-dimensional channel is in a topological setting from a situation in which it is not: with the application of a transverse electric field coupling to the dipole moment, the energies and corresponding velocities of all the plasmon modes would shift in the non-topological case, whereas in the topological case one of these modes will be insensitive to the electric field.

While the presence of an intrinsic dipole moment associated with plasmons in these one-dimensional systems is consistent with the presence of a QGD $\mathbfcal{D}$ in their two-dimensional realizations \cite{2021PlasmonQGD}, it is not necessary for $\mathbfcal{D} \ne 0$ for these one-dimensional plasmons to carry a transverse dipole moment.  Figs. \ref{fig:Intra energy K valley ES off 4 bands,m=0} and \ref{fig:Intra dipole K valley ES off 4 bands,m=0} illustrate this for the situation in which the gap parameter $m$ vanishes, so that $\mathbfcal{D} = 0$ for plasmons in this system in two dimensions \cite{2021PlasmonQGD}.  Clearly one finds a non-vanishing transverse dipole for such plasmons, and indeed the results are qualitatively similar to those found for $\lambda\lambda_0 = 1$.  Note that one does not expect the one-dimensional system to host edge states when $\Delta=2m=0$.
% numerical grid parameter $Ly =   640000$ {\AA},
%The parameters for WeS$_2$ is given in the main text,
%$\Delta =    1.6$ eV,$v_F\hbar =    3.939$ eV$\cdot${\AA},
%the Fermi energy is always $E_F = 1.00695$ eV.
\begin{figure*}[!htb]
	\centering
    \begin{adjustbox}{minipage=\linewidth,scale=1.0}
	\begin{subfigure}[b]{0.495\textwidth}
		\caption[]%
		{{}}
		\centering
		\includegraphics[width=\textwidth]{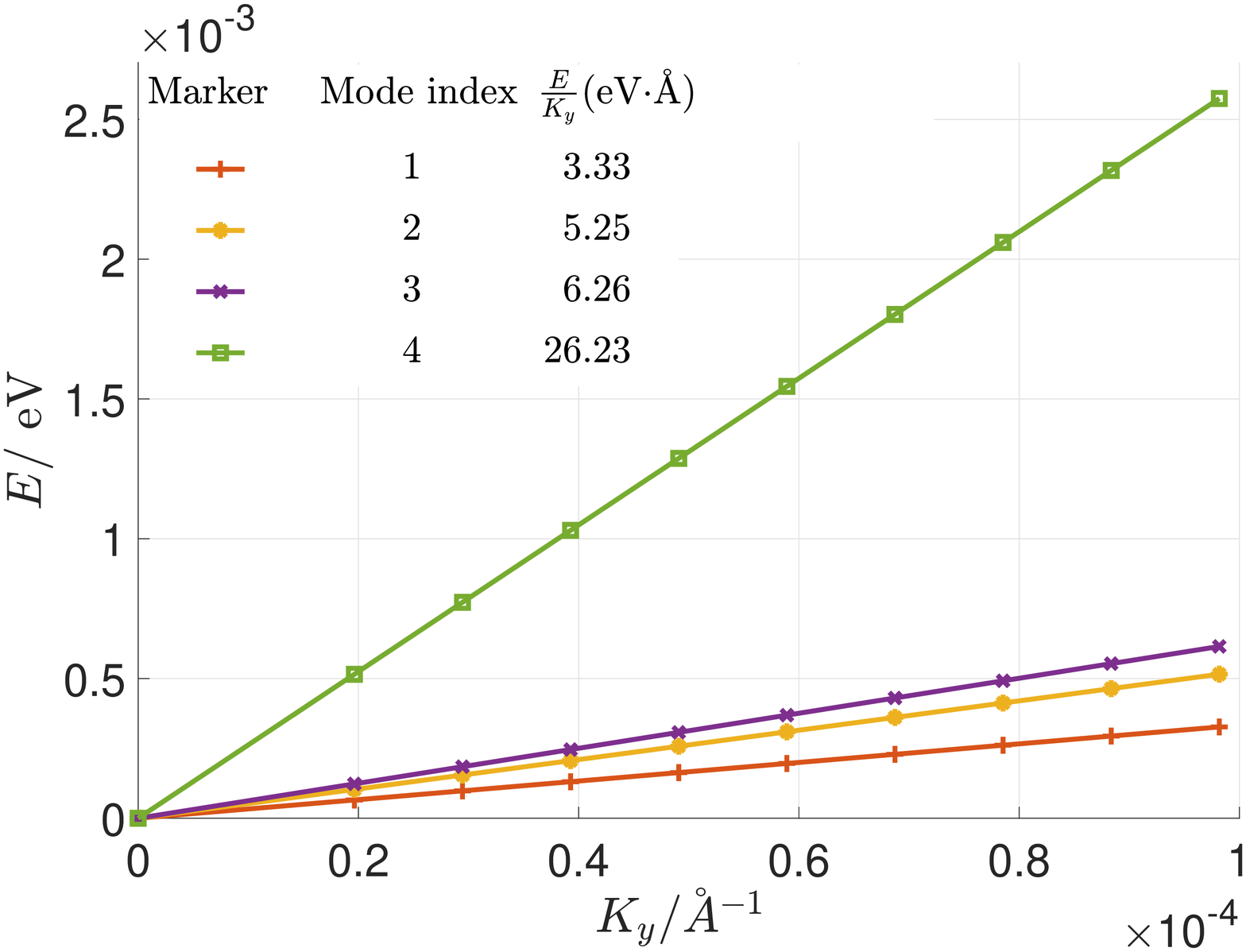}
		\label{fig:Intra energy K valley ES off 4 bands,m=0}
	\end{subfigure}
	\hfill
	\begin{subfigure}[b]{0.495\textwidth}
		\caption[]%
		{{}}
		\centering
		\includegraphics[width=\textwidth]{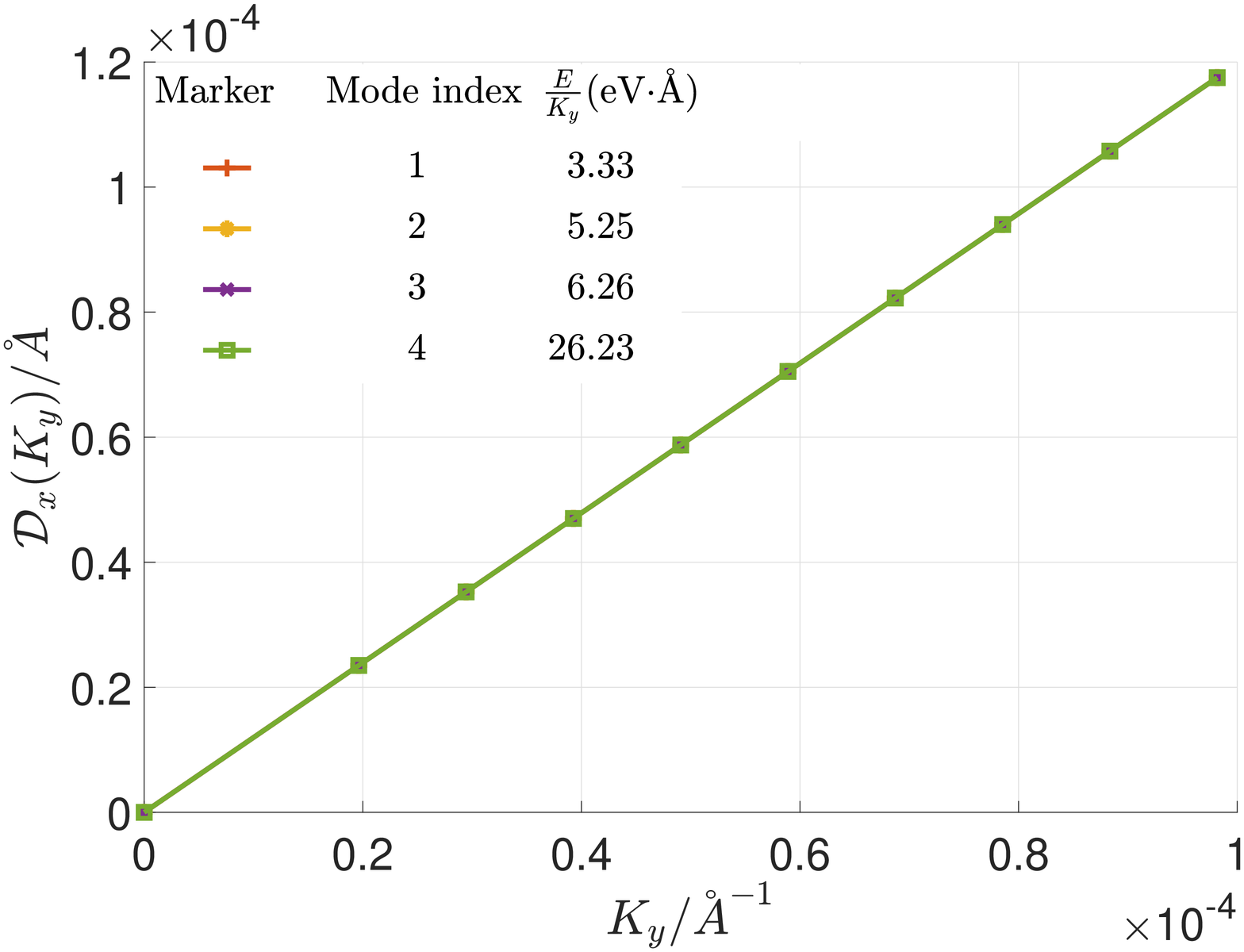}
		\label{fig:Intra dipole K valley ES off 4 bands,m=0}
	\end{subfigure}
    \end{adjustbox}
	\caption[]
	{\small
    Intraband plasmon energies and transverse dipole moments for $K$ valley excitations in a situation where 4 subbands are occupied, for a gapless chiral fermion.
    Parameters are $L =   50$ {\AA}, $\Delta =    0$ eV,$ v_F\hbar =    3.94$ eV$\cdot${\AA}, Fermi energy $\varepsilon_F = 1.007$ eV.
    (a) Plasmon energies.
    (b) Corresponding transverse dipole moments.
    }
	\label{fig:Intra energy and dipole K valley ES off 4 bands,m=0}
\end{figure*}

\begin{figure*}
	\centering
	\begin{adjustbox}{minipage=\linewidth,scale=1.0}
	\begin{subfigure}[b]{0.495\textwidth}
		\caption[]%
		{{}}
		\centering
		\includegraphics[width=\textwidth]{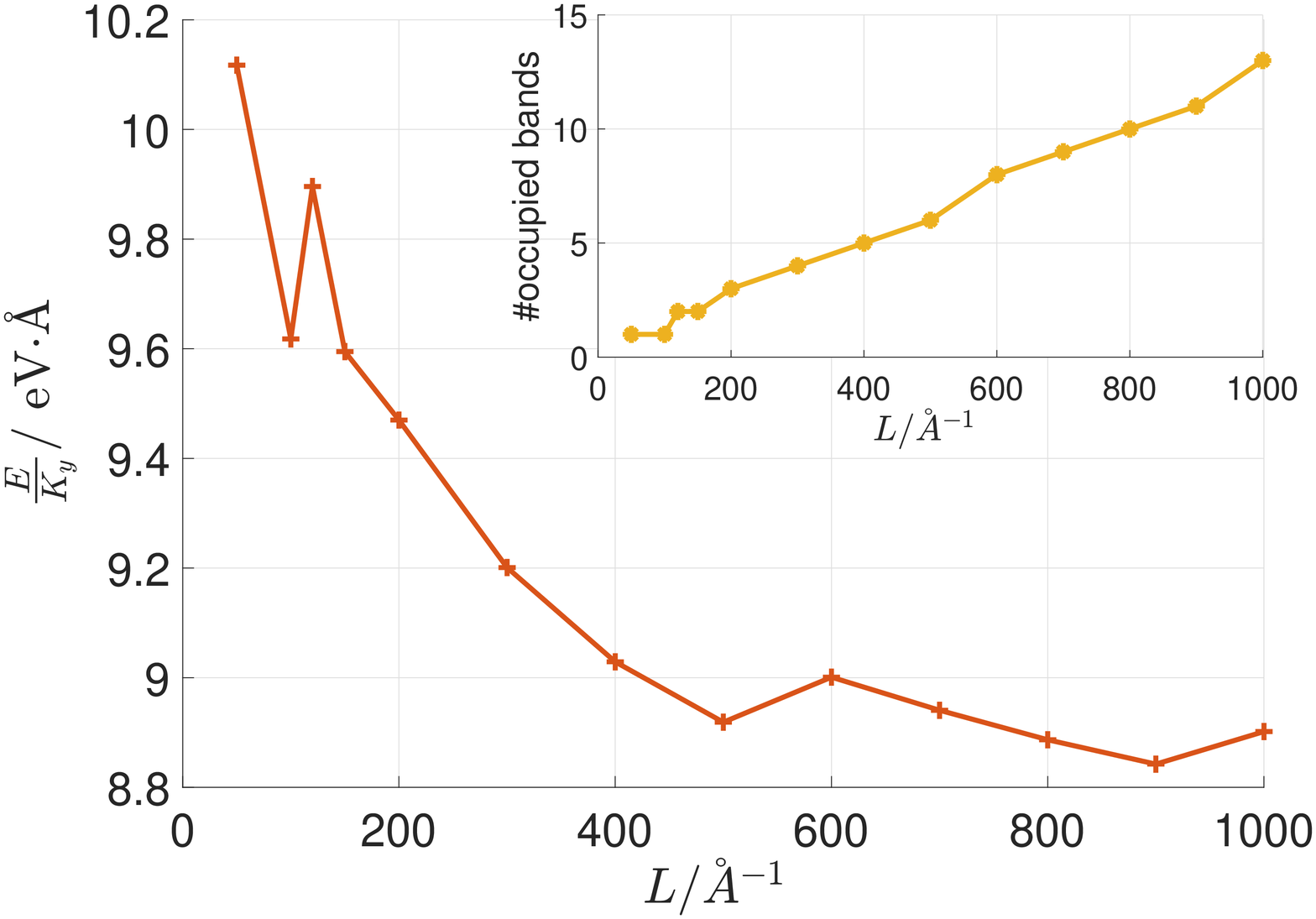}
		\label{fig:Intra energy K valley graphene vs L}
	\end{subfigure}
	\hfill
	\begin{subfigure}[b]{0.495\textwidth}
		\caption[]%
		{{}}
		\centering
		\includegraphics[width=\textwidth]{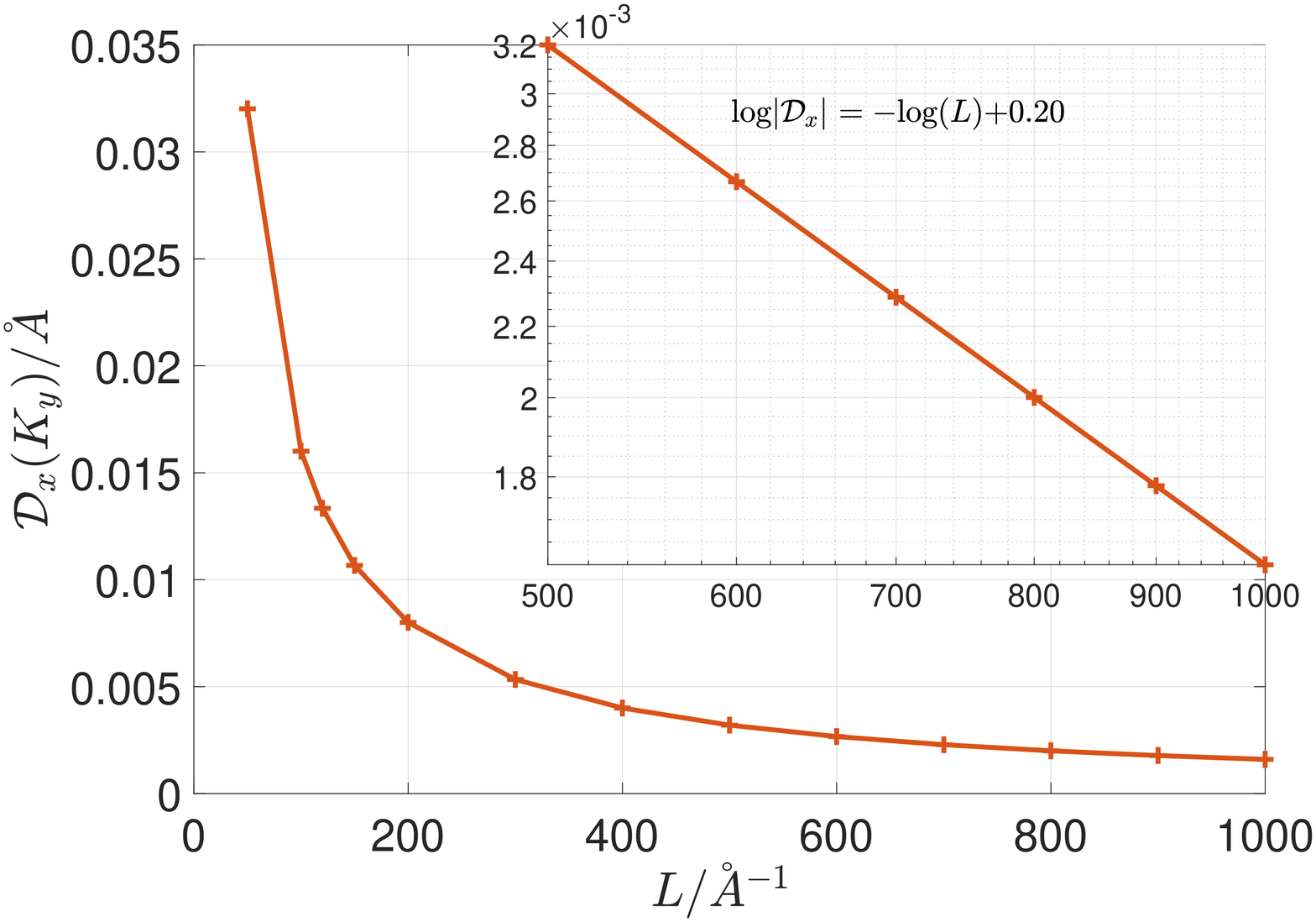}
		\label{fig:Intra dipole K valley graphene vs L}
	\end{subfigure}
	\vskip\baselineskip
	\begin{subfigure}[b]{0.495\textwidth}
		\caption[]%
		{{}}
		\centering
		\includegraphics[width=\textwidth]{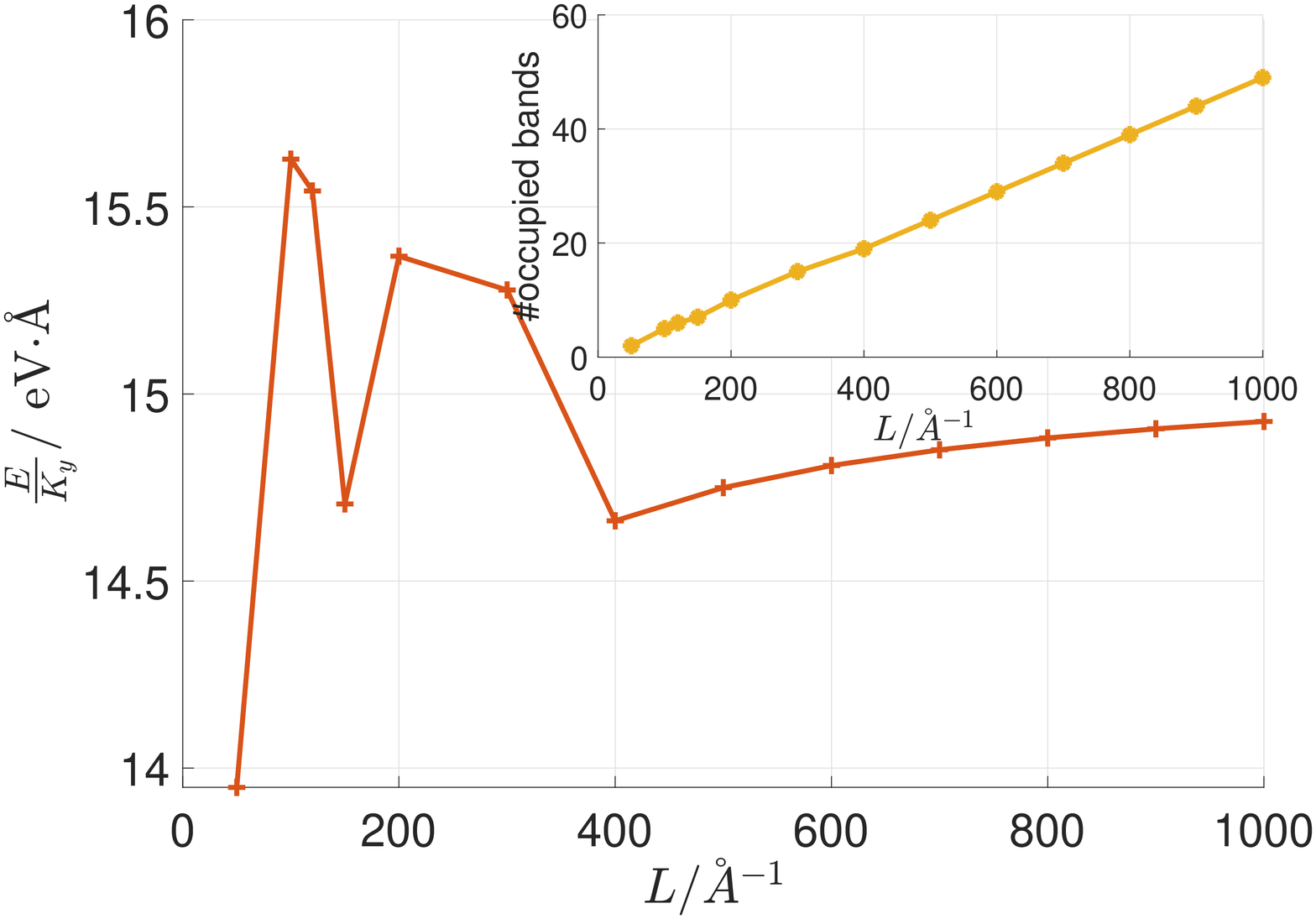}
		\label{fig:Intra energy K valley gapped vs L}
	\end{subfigure}
	\hfill
	\begin{subfigure}[b]{0.495\textwidth}
		\caption[]%
		{{}}
		\centering
		\includegraphics[width=\textwidth]{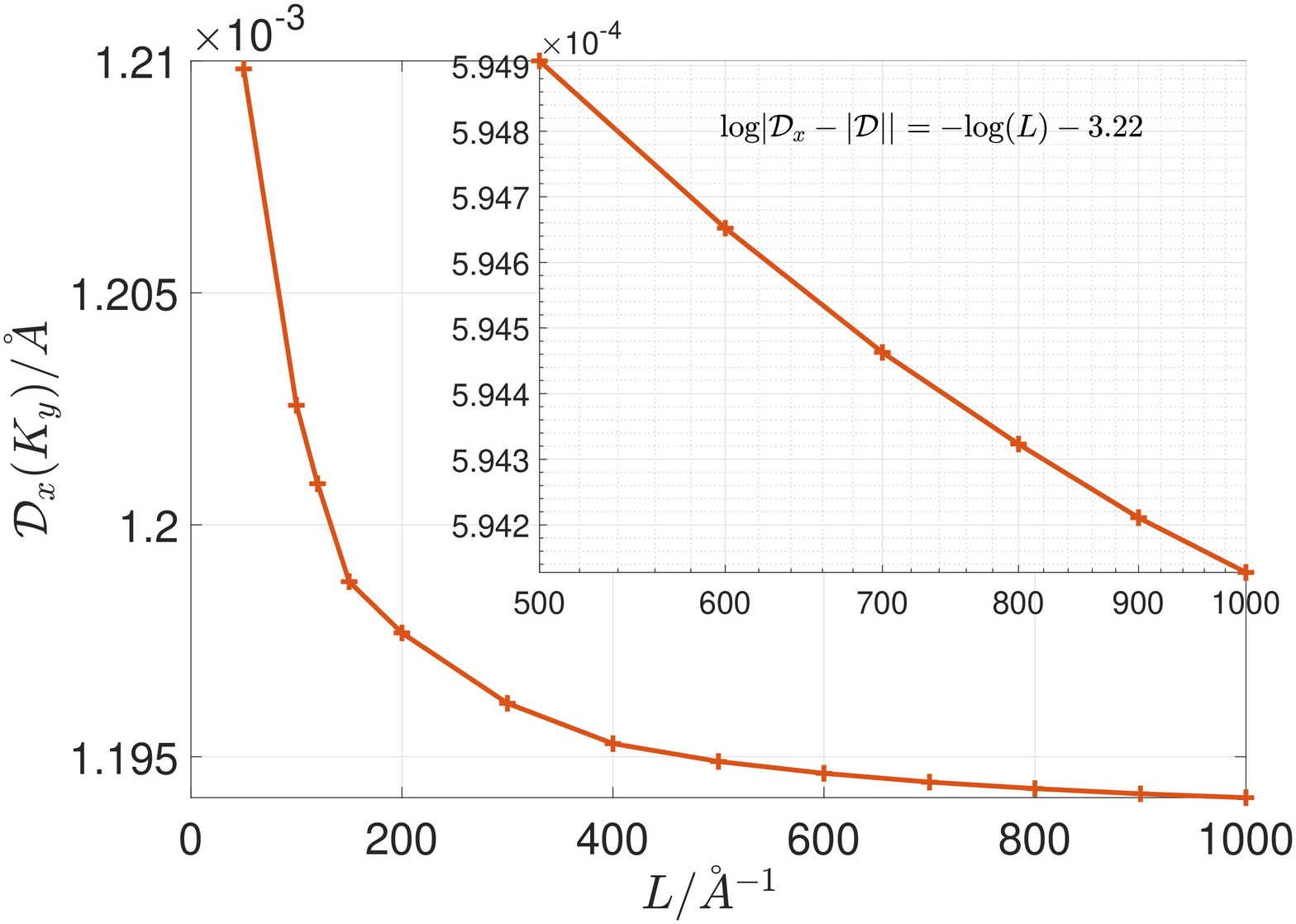}
		\label{fig:Intra dipole K valley gapped vs L}
	\end{subfigure}
	\end{adjustbox}
	\caption[]
	{ Single valley intraband plasmon velocity $\frac{E}{K_y}$ and transverse dipole moment evaluated at fixed plasmon momentum $K_y = 9.8\times 10^{-5} \text{\AA}^{-1}$.  Results are shown for the highest energy intraband plasmon.
    (a) Plasmon velocity for fixed Fermi energy 0.252 eV with zero gap. Horizontal axis shows the width of the wire. The energy gap is set to zero, so that two-dimensional quantum geometric dipole $\mathbfcal{D}$ vanishes. The cusps in the curve correspond to level anticrossings.
    (b) Transverse plasmon dipole moments corresponding to (a). Inset is a log-log plot of the same results. This quantity extrapolates to zero in this case.
    (c) Plasmon velocity for fixed Fermi energy 1.007 eV. Horizontal axis shows the width of the wire. The energy gap is that of the WSe$_2$, i.e. $\Delta = 2 m = 1.60$eV. The cusps in the curve corresponds to level anticrossings.
    (d) Transverse plasmon dipole moments corresponding to (c). Inset is a log-log plot of the same results. The $L \rightarrow \infty$ value of the dipole moment $|\mathbfcal{D}|\approx 6.00\times 10^{-4}$ for these system parameters. The limiting value is the same as the 2D QGD.
    }
	\label{fig:effective velocity and dipole vs L}
\end{figure*}

While for these relatively narrow systems we see little difference in the behavior of one-dimensional plasmons between systems in which $\mathbfcal{D} \ne 0$ and $\mathbfcal{D} =0$, the distinction becomes relevant as the conducting channel gets wider.
We illustrate this by computing the plasmon transverse dipole moment both for a chiral fermion system with vanishing gap ($m=0$), for which $\mathbfcal{D}=0$  \cite{2021PlasmonQGD}, and for a gapped chiral fermion, for which it does not, and examine the plasmon behavior as the width $L$ increases while the Fermi energy $\varepsilon_F$ is held fixed.  Fig. \ref{fig:Intra energy K valley graphene vs L} illustrates the behavior of the velocity of the fastest plasmon for a system with $m = 0$, and the associated dipole moment can be seen to vanish as $L$ becomes large (Fig.  \ref{fig:Intra dipole K valley graphene vs L}).  Figs. \ref{fig:Intra energy K valley gapped vs L} and \ref{fig:Intra dipole K valley gapped vs L} illustrate the corresponding quantities for a system with $m \ne 0$, for which the transverse dipole moment matches onto the (two-dimensional) quantum geometric dipole magnitude $|\mathbfcal{D}|$. The robustness of the plasmon dipole moment with increasing $L$ is thus a signature of its quantum geometric nature.

%The graphene ribbon though has zero effective mass has nonvanishing dipole when the ribbon has finite width, but when the ribbon goes to the 2D limit its dipole will vanish if we fix the Fermi energy and increase the width $L$, and it has a nearly $L^{-1}$ decay when $L \rightarrow \infty$(see the loglog inset on the right.)
%\FloatBarrier
%\FloatBarrier
%For gapped Dirac materials, $m\neq 0$, the dipole moment is nonvanishing even for the 2D limit ($L \rightarrow \infty$). Here we show the dipole and energy for the highest energy plasmon mode varies with ribbon width $L$ for a gapped graphene with gap $\Delta = 2 m = 1.60$eV, and fixed plasmon total momentum $K_y = 9.8\times 10^{-5} \text{\AA}^{-1}$.
%\FloatBarrier
%\FloatBarrier
%Its has been shown that for finite width, the single valleys plasmon carry a nonvanishing dipole moment. The effective mass of the material determines whether the dipole vanishes in the 2D limit.

\subsection{Mulitple Valleys: Vanishing Dipole for Time-Reversal Symmetric Systems}

\begin{figure*}
	\centering
	\begin{adjustbox}{minipage=\linewidth,scale=1.0}
	\begin{subfigure}[b]{0.495\textwidth}
		\caption[]%
		{{}}
		\centering
		\includegraphics[width=\textwidth]{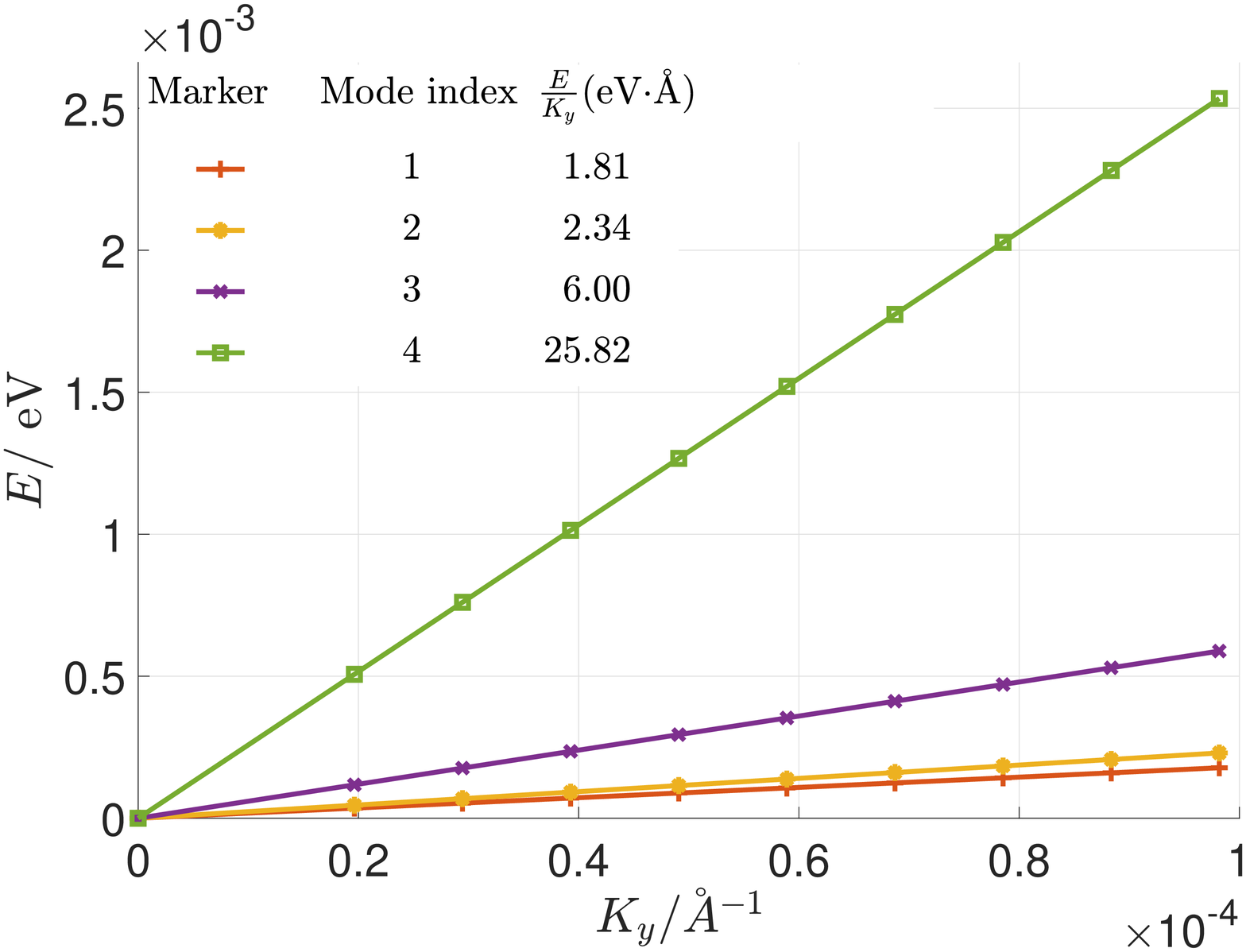}
		\label{fig:Intra energy both valley ES off 2 bands}
	\end{subfigure}
	\hfill
	\begin{subfigure}[b]{0.495\textwidth}
		\centering
		\caption[]%
		{{}}
		\includegraphics[width=\textwidth]{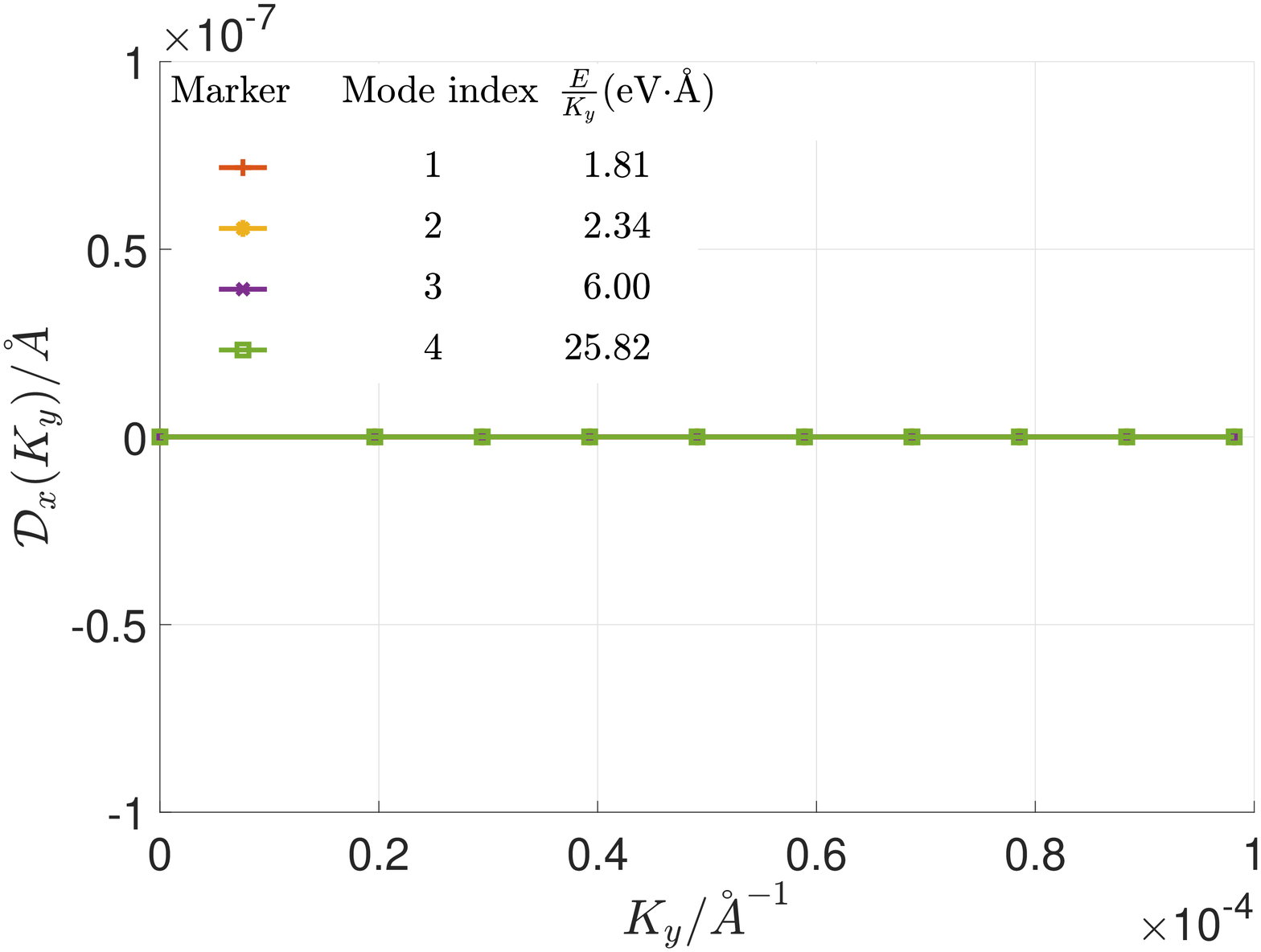}
		\label{fig:Intra dipole both valley ES off 2 bands}
	\end{subfigure}
	\vskip\baselineskip
	\begin{subfigure}[b]{0.495\textwidth}
		\centering
		\caption[]%
		{{}}
		\includegraphics[width=\textwidth]{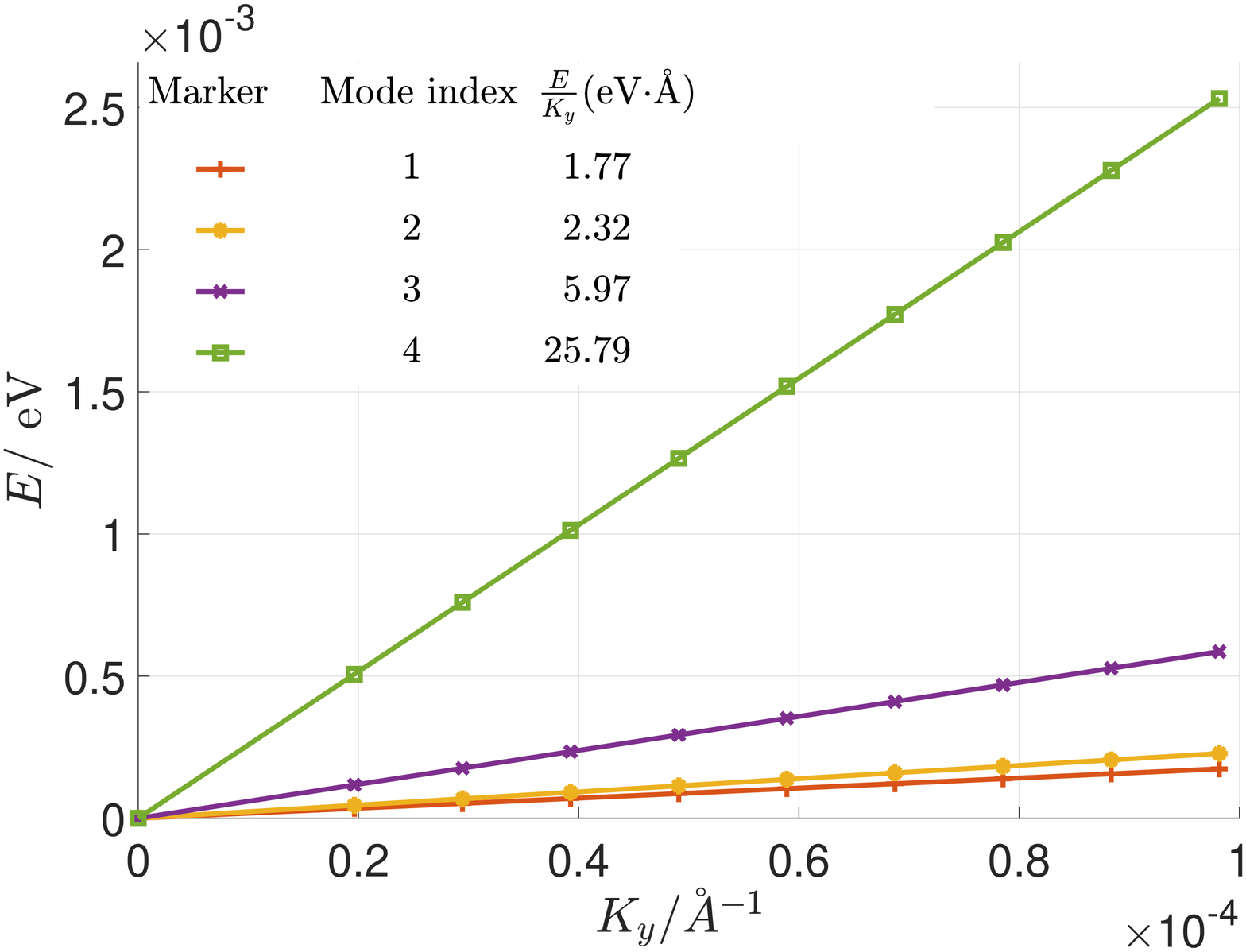}
		\label{fig:Intra energy imba valley(1,0.85) ES off 2+2 bands}
	\end{subfigure}
	\hfill
	\begin{subfigure}[b]{0.495\textwidth}
		\caption[]%
		{{}}
		\centering
		\includegraphics[width=\textwidth]{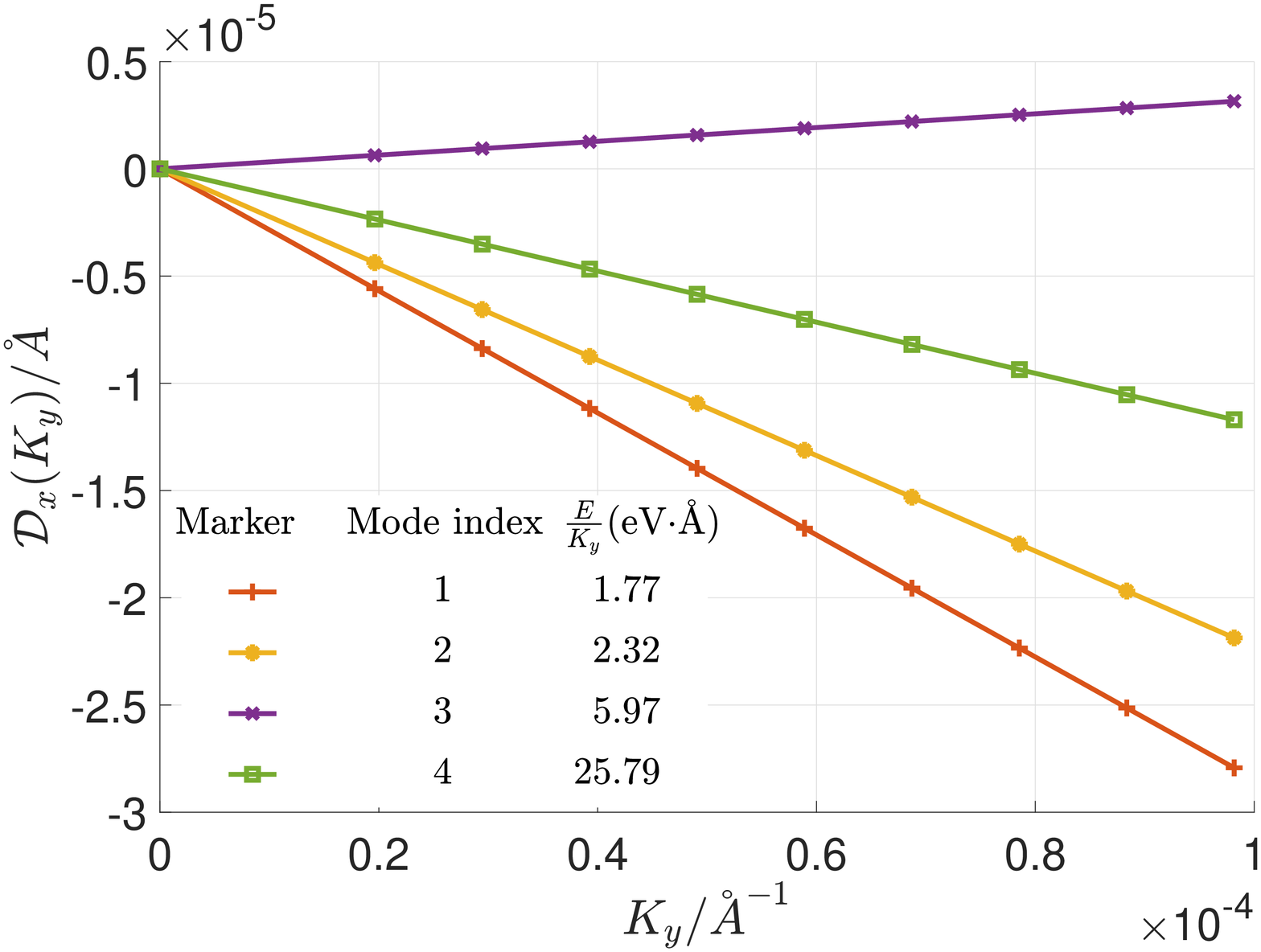}
		\label{fig:Intra dipole imba valley(1,0.85) ES off 2+2 bands}
	\end{subfigure}
	\end{adjustbox}
	\caption[]
	{ Intraband plasmon energies and associated transverse dipole moments when two valleys that are time-reversal partners are retained. $L =   50$ {\AA}, $\Delta =    1.6$ eV, $v_F\hbar =    3.94$ eV$\cdot${\AA}.
    (a) Plasmon energies when 2 subbands are occupied in each valley.  The Fermi energy $\varepsilon_F = 1.008$ eV.
    (b) Transverse dipole moment corresponding to (a).  Due to the symmetry, all the plasmon modes have vanishing dipole.
    (c) Plasmon energies for imbalanced valleys where 2 subbands are occupied in each valley. Fermi energy in $K$ valley is $\varepsilon_F(\tau=1) = 1.008$ eV, while in $K'$ valley $\varepsilon_F(\tau=-1) = 1.000$ eV.  The different effective Fermi energies can be induced by a magnetic field coupling to electron spins.
    (d) Transverse dipole moments corresponding to (c). Due to broken valley symmetry, all the plasmon modes yield nonvanishing dipole.
    }
	\label{fig: valley symmetry and imbalance for zero and nonzero dipole}
\end{figure*}

While the systems discussed above involve relatively simple Hamiltonians, their physical realizations require time-reversal symmetry-breaking, for example via ferromagnetic insulating films which would need to be patterned onto the surface of a topological insulator.  A much simpler system to realize would be a nanoribbon of transition metal dichalcogenide (TMD) material, which in most cases preserves time reversal symmetry.  Such systems typically have two valleys, which are time-reversal partners of one another.  With time-reversal symmetry intact, one does not expect plasmon modes to carry an intrinsic dipole moment.  Figs. \ref{fig:Intra energy both valley ES off 2 bands} and \ref{fig:Intra dipole both valley ES off 2 bands} illustrate such a situation.
%%\FloatBarrier
%\captionsetup{position=top}

Non-vanishing dipole moments in such systems can be induced by breaking the symmetry between valleys. In TMD materials, spin-orbit coupling leads to a splitting between spin-up and spin-down states in opposite directions for the two valleys \cite{Xiao_2012}, so that a magnetic field imbalances the populations of the valleys through the Zeeman coupling.  For magnetic fields that are not too large, such that the magnetic length $\ell=\sqrt{\hbar c/eB}$, with $B$ the magnetic field, is larger than the ribbon width, the orbital motion of the electrons will largely be unaffected by the field.  To a good approximation one then only needs to include the spin dependence of the Fermi surfaces to account for the field.

\begin{figure*}
	\centering
	\begin{adjustbox}{minipage=\linewidth,scale=1.0}
	\begin{subfigure}[b]{0.495\textwidth}
		\caption[]%
		{{}}
		\centering
		\includegraphics[width=\textwidth]{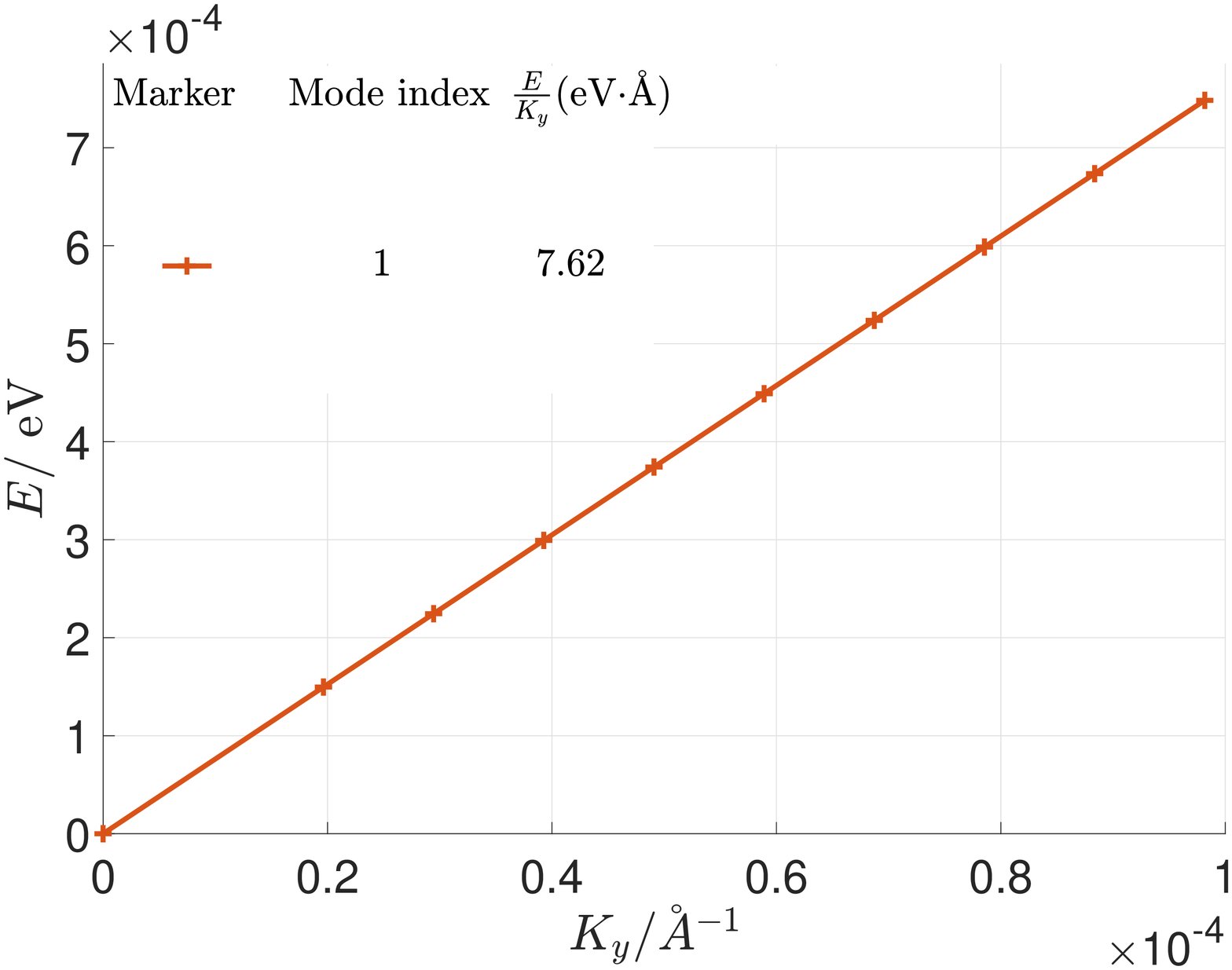}
		\label{fig:WSe2 Intra energy K' valley Depletion}
	\end{subfigure}
	\hfill
	\begin{subfigure}[b]{0.495\textwidth}
		\caption[]%
		{{}}
		\centering
		\includegraphics[width=\textwidth]{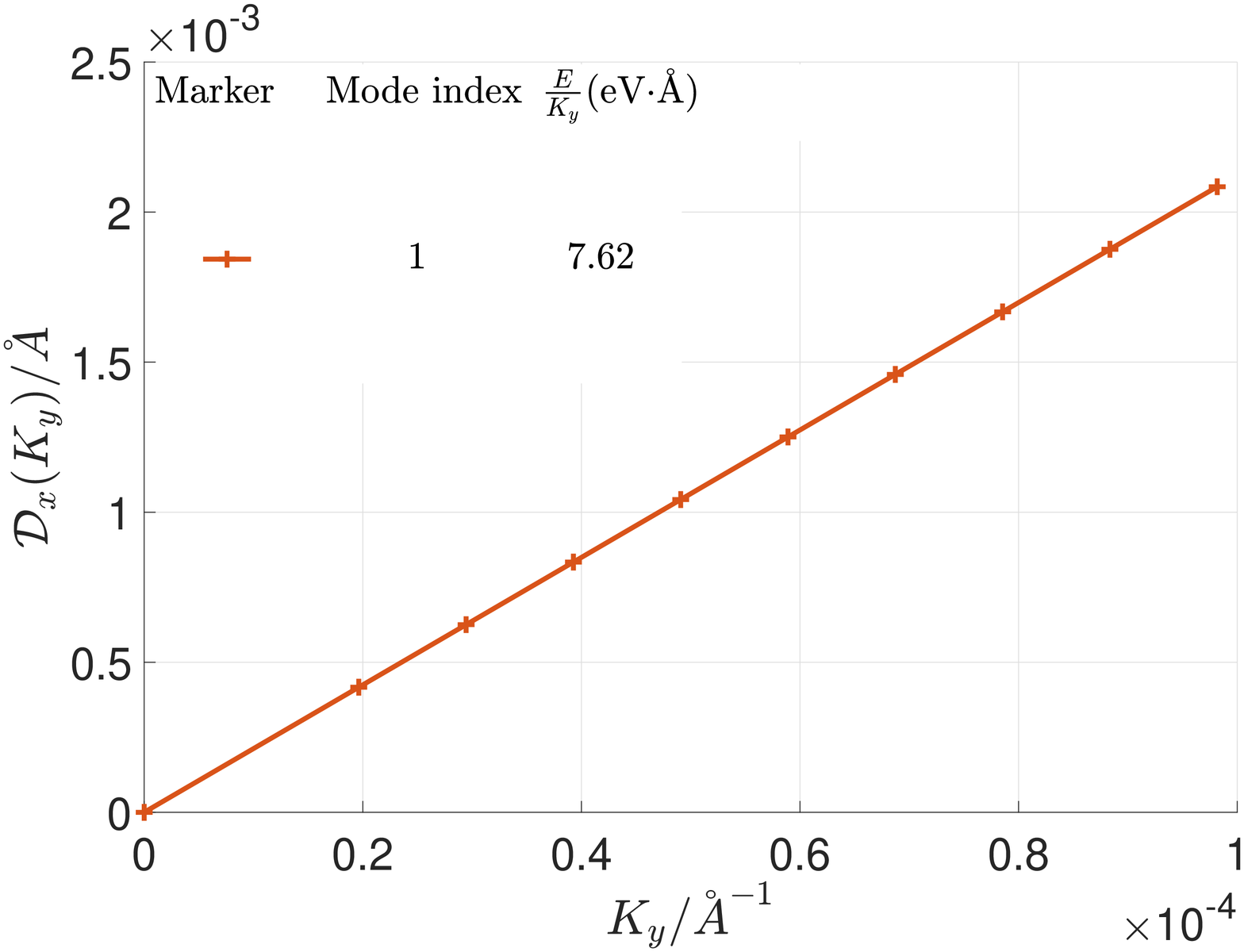}
		\label{fig:WSe2 Intra dipole K' valley Depletion}
	\end{subfigure}
	\end{adjustbox}
	\caption[]
	{
    Plasmon energies and transverse dipole moments in a 50 \AA wire with Fermi energy $\varepsilon_F=$-0.838 eV, for a typical TMD (WSe$_2$) half gap of 0.8 eV, and Zeeman splitting of 8 meV. The boundary condition is chosen so that their are no edge modes. Only one subband of one valley is occupied in this situation.
    (a) Plasmon energies.
    (b) Transverse dipole moments.
    }
	\label{fig:WSe2 Intraband K' valley Depletion}
\end{figure*}

%Figs. \ref{fig:Intra energy both valley ES off 2 bands} and \ref{fig:Intra dipole both valley ES off 2 bands} illustrate this situation in the absence of a magnetic field.  In this case each valley has two occupied subbands and the four resulting plasmon modes above the particle-hole continuum are apparent.  To within numerical accuracy, the dipole moments of all of them vanish, as expected.  On the other hand one may consider situations in which the occupied bands of the two valleys have opposite spins, as is the case for the highest hole bands of TMD materials \cite{Xiao_2012}.
Figs. \ref{fig:Intra energy imba valley(1,0.85) ES off 2+2 bands} and \ref{fig:Intra dipole imba valley(1,0.85) ES off 2+2 bands} illustrate such a situation in a ribbon of width $L=50$\AA , for system parameters modeled after the hole bands of WSe$_2$, with a magnetic field $B=10$T, for which $\ell \approx 81$\AA.  While the bands for each valley have the same dispersion, the Zeeman coupling leads to a difference of $\sim 8$meV in the extrema of the $K$ and $K'$ bands, yielding different carrier populations in each of them.  While this leads to little change in the plasmon dispersions (compare Figs. \ref{fig:Intra energy both valley ES off 2 bands} and \ref{fig:Intra energy imba valley(1,0.85) ES off 2+2 bands}), the plasmons now carry dipole moments with small magnitudes (Fig. \ref{fig:Intra dipole imba valley(1,0.85) ES off 2+2 bands}.)  Interestingly, these dipole moments can have either sign.  At the more extreme end, this effect can used to completely depopulate one of the valleys of carriers.  This situation is illustrated in Figs. \ref{fig:WSe2 Intra energy K' valley Depletion} and \ref{fig:WSe2 Intra dipole K' valley Depletion}.  Interestingly this yields a plasmon dipole moment that is relatively large.

%However, if the two valleys are not equally populated, we are able to generate a non vanishing dipole moment.
%Here we show an illustrative example of imbalanced valley population by artificially set different Fermi energies in K and K' Valleys.
%The K and K' valley Fermi energies are 1.00695eV and 0.99895eV respeactively. Note that 8 meV is the energy scale for a typical zeeman splitting achievable for WSe2 at 10 T magnetic field.
%\subsection{Intraband Plasmons: Imbalanced Valleys including Edge state}

%For topological insulator, the sign of the Chern number in the vacuum's K valley is the opposite of that in the ribbon's K valley and the sign of Chern number in K' valleys is the same as that in the ribbon, or vice versa. Valley symmetry is broken naturally in 2 ways. First, only one valley will have edge mode and the other will not. Second, the quantized wave vectors will not take the same value, because of the $ \lambda_0, \tau$ dependence in \eqref{eq: bulk state kx quantization},
%$$k_x(n,\tau=1,\lambda_0=-1) \neq k_x(n,\tau=-1,\lambda_0=-1).$$
%As a result, non vanishing dipole can be expected from broken valleys symmetries.

\section{Summary and Discussion}
In some two-dimensional conducting systems, plasmon excitations come with an intrinsic dipole moment that is quantum geometric in nature \cite{2021PlasmonQGD}.
In this work, we have explored conditions under which this kind of dipole moment might be found for quasi-one-dimensional geometries of the corresponding systems, using an RPA approach. Our studies focused on chiral fermions, as might be found on the surface of a topological insulator, or in two-dimensional van der Waals materials such as graphene or TMD's, and we adopted a simplified model with infinite mass boundary conditions at the system edges.
We found that the presence of a transverse dipole moment is more ubiquitous in the wire geometry than in the two-dimensional system: the opening of a gap in the spectrum due to transverse confinement is sufficient to stabilize it, even when the corresponding gapless spectrum has no such dipole moment in two dimensions.  The connection with the quantum geometric dipole is made by considering the wide wire limit, for which the plasmons retain dipole moments when the corresponding two dimensional system has a non vanishing quantum geometric dipole.

{These plasmons differ from the corresponding modes of more conventional semiconducting quantum wires (e.g., GaAs): the topological character of the chiral fermions allow the possibility that they support edge states, which we find are present for sufficiently wide wires.  Their presence increases the number of gapless plasmon modes supported by the systems, but they only differ in their qualitative behavior for systems with relatively large gaps.  In this limit we found that the dipole moments of the plasmon modes supported by the transverse confined states became degenerate across the modes, whereas a single plasmon associated with the edge states has vanishing dipole moment.}

Single chiral fermion flavors can only be found in systems with broken time-reversal symmetry.  In principle a conducting channel of these could be fabricated on the surface of topological insulator with ferromagnetically ordered magnetic ions on it surface, with a one-dimensional channel cut through.  Systems with both chiral fermions and time-reversal symmetry support multiple flavors of chiral fermions (valleys), which are time-reversal partners of one another.  Plasmons in these systems do not carry dipole moments.  However in TMD's they can be induced by the introduction of a magnetic field, which due to spin-orbit coupling imbalances the carrier populations in different valleys.  For low carrier densities one may find complete depletion of some valleys, leading to relatively large transverse dipole moments in the plasmons.

A transverse dipole moment associated with a plasmon in a nanowire will in principle be signaled by a sensitivity of its frequency and speed to the application of a transverse electric field \cite{Pizzochero_2021}: these will both vary linearly with electric field.  For example, for the system parameters considered in
Figs. \ref{fig:WSe2 Intra energy K' valley Depletion} and \ref{fig:WSe2 Intra dipole K' valley Depletion},
an external electric field of 0.01 V$\cdot$\AA$^{-1}$ will modify the speed of the plasmon by approximately $3.2\times 10^6$m/s, and its energy by approximately 3\%.  Control of plasmon energies in such a continuous way in a single system could in principle bring new capabilities to plasmonic systems incorporating such nanowires.

Our studies suggest further directions for exploration.  Beyond the gapless plasmons we have focused on in this work, nanowires support gapped, intersubband plasmons at higher frequencies \cite{1989DasSarma}.  Preliminary work \cite{unpublished} indicates that these also obtain non-vanishing dipole moments in chiral fermion settings, and they offer a way to detect this physics at higher frequencies.  One may also consider the presence and role of transverse dipole moments in more complicated settings than considered in this work.
For example, understanding the behavior of nanowire plasmons for other classes of boundary conditions, in particular those for which valley mixing is induced at the single-particle level \cite{Brey_2006,2008Akhmerov}, would likely be important for many types of nanowires.  One may also consider the effects of transverse dipole moments on plasmon nanochannel networks, which arise naturally in moir{\'e} superlattices, and which have been shown to support their own unique dynamics \cite{Ni_2015,Brey:2020aa,Stauber:2020aa}.  Such systems under some circumstances may become spontaneously valley-polarized, opening another avenue for the broken time-reversal symmetry needed for transverse dipole moments.

Plasmon modes can be generally understood as collective oscillations of the electric dipole moments of conducting systems.  That such oscillations can occur with a static component which depends on the plasmon momentum is a relatively new insight, allowing the possibility of new physical phenomena.  Nanowires offer a setting in which the direction of these dipole moments are fixed, so that they can be interrogated with electric fields.  For systems that support them, we expect they will admit new plasmonic phenomomena of fundamental interest.

\section*{Acknowledgements}
L.B. acknowledges funding from PGC2018-097018-B-I00 (MICIN/AEI/FEDER, EU).
HAF and JC acknowledge the support of the NSF  through Grant Nos. DMR-1914451 and ECCS-1936406.
HAF acknowledges support from the US-Israel
Binational Science Foundation (Grant Nos. 2016130 and 2018726),
of the Research Corporation for Science Advancement through a Cottrell SEED Award.

\begin{widetext}

\newpage
\section*{Appendix A: Nanowire with Infinite Mass Boundary Condition}

In order to model chiral Dirac fermions confined to a quasi-one-dimensional channel, we consider a two-dimensional system with a position-dependent mass term,
\begin{equation*}
%\label{eq: ribbon mass term}
  \tilde{m}(x) =
    \begin{cases}
    m,   &0<x<L , \\
    V_0, &x<0 \text{ or } x>L.
    \end{cases}
\end{equation*}
The Hamiltonian of the system is, in general,
\begin{equation*}
%\label{eq:Hamiltonian}
 H=\left(
   \begin{array}{cc}
     \tilde{m} & i\tau k_x  - i k_y\\
     i\tau k_x + i k_y & -\tilde{m} \\
   \end{array}
 \right),
 \end{equation*}
where $\tau$ is the valley index.
We find stationary states by matching eigenstates of $H$ in the regions $-\infty < x <0$ (region $I$), $0 \le x \le L$ (region $II$) and $x > L$ (region $III$) at the locations $x=0$ and $x=L$.  We ultimately take the limit $V_0 \rightarrow \pm\infty$ (infinite mass boundary conditions \cite{1987Berry}), but some care must be taken with regard to whether the Chern numbers in the central and outer regions are the same or different.  These Chern numbers are given by $C=\tau\,{\rm sgn}(\tilde{m})/2$, so that the two cases are distinguished by the sign of $mV_0$.  For the regions outside the nanowire, we denote the sign of the Chern numbers by $\lambda_0 \equiv {\rm sgn} (\tau V_0)$.  Outside the wire one has for region I
\begin{equation*}
%\label{eq: region I evanescent solution}
  \psi^{I}(\vec{r}) = C_1
\left(
  \begin{array}{c}
    V_0 + E \\
    i(-\tau_0 k_1+k_y) \\
  \end{array}
\right)
   e^{k_1 x + i k_y y },
\end{equation*}
with $k_1^2 = V_0^2- E^2+ k_y^2,$ where $E$ is the energy of the state, while in region III,
\begin{equation*}
%\label{eq: region III evanescent solution}
  \psi^{III}(\vec{r}) = C_3
\left(
  \begin{array}{c}
    V_0 + E\\
    i(\tau_0 k_3+k_y) \\
  \end{array}
\right)
   e^{-k_3 x + i k_y y },
\end{equation*}
with
$k_3^2 = V_0^2- E^2+ k_y^2.$  Writing the upper and lower components of the spinors as $\psi_1$ and $\psi_2$, respectively, with the limit $|V_0| \rightarrow \infty$ we obtain \cite{1987Berry}
\begin{equation}
\label{eq: infinte mass BC ribbon}
  \left.\frac{\psi_1}{\psi_2}\right|_{x=0} = i \lambda_0, \quad \left.\frac{\psi_1}{\psi_2}\right|_{x=L} = -i \lambda_0.
\end{equation}

In the interior of the wire (region II), the general form of the wavefunction is
\begin{equation}
\label{eq: ribbon state def}
  \psi_{\vec{k}}(\vec{r}) = \frac{e^{i k_y y}}{\sqrt{2 E( E+m)}}
  \left[ A
  \left(
    \begin{array}{c}
      E+m \\
      \tau k_x + i k_y \\
    \end{array}
  \right) e^{i k_x x}
  + B
  \left(
    \begin{array}{c}
      E+m \\
      -\tau k_x + i k_y \\
    \end{array}
  \right)e^{- i k_x x}
  \right]
\end{equation}
with $A$ and $B$ constants to be determined.  Using Eqs. \ref{eq: infinte mass BC ribbon} one finds
\begin{equation}
\label{eq: bulk x=0 simp 2}
  \frac{A}{B} = - \frac{ E+m - i \lambda_0 (-\tau k_x + i k_y) }{ E+m - i \lambda_0 (\tau k_x + i k_y) }
\end{equation}
and
\begin{equation}\label{eq: bulk x=L simp 2}
  \frac{A}{B} = - e^{-2i k_x L} \frac{ E+m + i \lambda_0 (-\tau k_x + i k_y) }{ E+m + i \lambda_0 (\tau k_x + i k_y) }.
\end{equation}
These two equations are consistent provided
\begin{equation*}
%\label{eq: bulk state kx condition 0}
  e^{2i k_x L} = \frac{m - i \lambda_0 \tau k_x}{m + i \lambda_0 \tau k_x} = e^{-2i  \lambda_0 \tau  \arctan (\frac{k_x}{m})},
\end{equation*}
or equivalently,
\begin{equation*}
%\label{eq: bulk state kx condition}
  k_x L =- \lambda_0 \tau  \arctan (\frac{k_x}{m})+ n \pi.
\end{equation*}
Using Eqs. \ref{eq: bulk x=0 simp 2} and  \ref{eq: bulk x=L simp 2}, one obtains the expressions
\begin{equation*}
%\label{eq: A bulk state}
  A \equiv A(\tau; \vec{k}) = \tau N \sqrt{(E+m)^2+ (\tau k_x-i k_y)^2} \sqrt{m + i \lambda_0 \tau k_x},
\end{equation*}
and
\begin{equation*}
%\label{eq: B bulk state}
  B \equiv B(\tau; \vec{k}) = -\tau N \sqrt{(E+m)^2+ (\tau k_x+i k_y)^2} \sqrt{m - i \lambda_0 \tau k_x},
\end{equation*}
where $N$ is a normalization constant,
\begin{equation*}
%\label{eq: bluk N}
  N =\left\{ 8 L_y E(E+m)^2 \left[ L (m^2+ k_x^2)+  \lambda_0 \tau m  \right]\right\}^{-1/2},
\end{equation*}
with $L_y$ the length of the one-dimensional region.

In addition to these confined states, the wire edges may support edge states.  This can occur only if the Chern numbers of the interior and exterior regions have different Chern numbers, which occurs when
$$\lambda\lambda_0=-1,$$
with $\lambda= \tau \,{\rm sgn}(m)$.  One finds these states by considering states that are evanescent not just in regions $I$ and $III$ but also in region $II$.  These have the form
\begin{equation*}
%\label{eq: wave function region II}
  \psi^{II}(\vec{r}) = A
  \left(
  \begin{array}{c}
      m+E \\
    i(\tau k_x + k_y) \\
  \end{array}
\right)
   e^{-k_x x + i k_y y }
   + B
\left(
  \begin{array}{c}
    m+E \\
    i(-\tau k_x+k_y) \\
  \end{array}
\right)
   e^{k_x x + i k_y y },
\end{equation*}
where here $k_x^2=m^2-E^2+k_y^2$.
Applying Eqs. \ref{eq: infinte mass BC ribbon} gives
\begin{align}
\frac{A(m+E) + B (m+E)} {A(\tau k_x+ k_y) - B (\tau k_x-k_y) }& = -\lambda_0 ,
\label{eq: infinite mass BC ribbon 1}\\
\frac{A(m+E)e^{-k_x L} + B (m+E)e^{k_x L} } {A(\tau k_x+k_y)e^{-k_x L} - B (\tau k_x-k_y)e^{k_x L} } & = \lambda_0 .
\label{eq: infinite mass BC ribbon 2}
\end{align}
Eliminating $A$ and $B$ in eqs. \eqref{eq: infinite mass BC ribbon 1} and \eqref{eq: infinite mass BC ribbon 2}, we arrive at a transcendental equation for $k_x$,
\begin{equation}
\label{eq: infinite mass BC ribbon 3, simp 4}
  e^{-2 k_x L} = \frac{|m|+\lambda_0 \lambda k_x}{|m| - \lambda_0 \lambda k_x}.
\end{equation}
Eq. \ref{eq: infinite mass BC ribbon 3, simp 4} may be solved if and only if
$ \lambda_0 \lambda =-1$
and $mL>1$.  These are the conditions under which the quasi-one-dimensional system hosts edge states.
Using Eqs. \ref{eq: infinite mass BC ribbon 1} and \ref{eq: infinite mass BC ribbon 2}, the explicit forms for the coefficients turn out to be
\begin{equation*}
%\label{def: A0, explicit}
  A \equiv A_0(\tau) = \tau \sqrt{\frac{ ( (m + E)^2 - (\tau k_x-k_y)^2 ) (m+ k_x) } {8E(E+m)^2 L_y\left( m  - L (m^2- k_x^2) \right)}},
\end{equation*}
and
\begin{equation*}
%\label{def: B0, explicit}
  B \equiv B_0(\tau) = - \tau
        \sqrt{\frac{ ( (m + E)^2 - (\tau k_x+k_y)^2 ) (m- k_x) } {8E(E+m)^2 L_y\left( m  - L (m^2- k_x^2) \right)}}.
\end{equation*}

\section*{Appendix B: Analytical Solution for Intraband Plasmons at Small Momentum}

The appearance of multiple plasmon modes may appear surprising.  In this section we show that this is to be expected, given the structure of the transverse wavefunctions for the systems we consider.  To do this we first develop an alternative, equivalent formalism by which one may find the plasmon excitations, and then apply it to the simple situation of two occupied subbands to show that there is more than a single gapless plasmon mode.

\subsection{Equivalent Dielectric Formalism}
For simplicity, we consider a massless chiral fermion system ($m=0$), which is equivalent to a single valley of graphene, for which we set $\tau=1$.
First, we start from eq. \eqref{eq: ribbon RPA eigen eq, bv}, and obtain the equivalent dielectric formalism for calculating plasmon frequency.  The contact interaction matrix element may be written in the form
\begin{align*}
%\label{eq: V contact interaction matrix}
 V_{n_1,n_2,n_3,n_4}(k_{y1},k_{y2},k_{y3},k_{y4})  = \frac{u_0}{2} L_y \int_0^{L} dx
 & \left[(\vec{A}^{\dagger}_{n_1,{k}_{y1}} e^{-ik_{x}(n_1) x} + \vec{B}^{\dagger}_{n_1,{k}_{y1}}
 e^{i k_{x}(n_1) x}) \cdot (\vec{A}_{n_4,k_{y4}} e^{ik_{x}(n_4) x} + \vec{B}_{n_4,k_{y4}} e^{-i k_{x}(n_4) x}) \right] \nonumber\\
 \times &
 \left[(\vec{A}^{\dagger}_{n_2,k_{y2}} e^{-ik_{x}(n_2) x} + \vec{B}^{\dagger}_{n_2,k_{y2}} e^{i k_{x}(n_2) x}) \cdot (\vec{A}_{n_3,k_{y3}} e^{ik_{x}(n_3) x} + \vec{B}_{n_3,k_{y3}} e^{-i k_{x}(n_3) x}) \right],
\end{align*}
where the spinor coefficients $\vec{A}_{n,k_y}$ and $\vec{B}_{n,k_y}$ may be read off from
Eq. \ref{eq: ribbon state def}, and $k_x(n)$ are the quantized wavevectors defined in Appendix A.
By defining
\begin{equation*}
%\label{def: f integral}
  f_{n_1,n_2,n_3,n_4}^{\sigma_1,\sigma_2,\sigma_3,\sigma_4} \equiv \int_0^L dx e^{-i \sigma_1 k_x(n_1)x - i \sigma_2 k_x(n_2)x + i \sigma_3 k_x(n_3)x + i \sigma_4 k_x(n_4)x},
\end{equation*}
and
\begin{equation*}
%\label{def: spinnor profile}
  \vec{D}_{n,k_y,\sigma} \equiv
  \Biggl\{
    \begin{array}{cc}
      \vec{A}_{n,k_y}, & \sigma = 1, \\
      \vec{B}_{n,k_y}, & \sigma = -1, \\
    \end{array}
%  \right\},
\end{equation*}
the interaction matrix elements can be written in the compact form
\begin{equation*}
%\label{eq: eq: V matrix alternative form}
   V_{n_1,n_2,n_3,n_4}(k_{y1},k_{y2},k_{y3},k_{y4})   = \frac{u_0}{2} L_y \sum_{\sigma_1,\sigma_2,\sigma_3,\sigma_4}
   \left[ \vec{D}_{n_1,k_{y1},\sigma_1}^{\dagger} \cdot \vec{D}_{n_4,k_{y4},\sigma_4}\right] \left[ \vec{D}_{n_2,k_{y2},\sigma_2}^{\dagger} \cdot \vec{D}_{n_3,k_{y3},\sigma_3}\right] f_{n_1,n_2,n_3,n_4}^{\sigma_1,\sigma_2,\sigma_3,\sigma_4}.
\end{equation*}

In the single valley case, the self-consistent equation for the plasmon wavefunction (Eq. \eqref{eq: ribbon RPA eigen eq, bv}) can now be written as
\begin{align*}
%\label{eq: ribbon RPA eigen eq rewrite}
a_{m_1,m_2,k_y'}(K_y) =& \frac{u_0 L_y}{\omega_{k_y} - \varepsilon(m_1,k_y'+K_y) + \varepsilon(m_2,k_y') } \nonumber\\
    &\times
    \sum_{n_2,n_3,k_{y1}}
    \sum_{\sigma_1,\sigma_2,\sigma_3,\sigma_4}
    \left[ \vec{D}_{m_1,k_{y}'+K_y,\sigma_1}^{\dagger} \cdot \vec{D}_{m_2,k_{y}',\sigma_4}\right] \left[ \vec{D}_{n_2,k_{y1},\sigma_2}^{\dagger} \cdot \vec{D}_{n_3,k_{y1}+K_y,\sigma_3}\right] f_{m_1,n_2,n_3,m_2}^{\sigma_1,\sigma_2,\sigma_3,\sigma_4} \nonumber\\
    &\times
    [n_F(n_2,k_{y1})  - n_F (n_3, k_{y1} + K_y) ] a_{n3,n_2,k_{y1}}(K_y)
\end{align*}

Defining
\begin{equation*}
%\label{def chi m1m2 s1s4}
  \chi_{m_1,m_2}^{\sigma_1,\sigma_4} (K_y) \equiv L_y \sum_{n_2,n_3,k_{y1}}\sum_{\sigma_2,\sigma_3}
  f_{m_1,n_2,n_3,m_2}^{\sigma_1,\sigma_2,\sigma_3,\sigma_4} \left[ \vec{D}_{n_2,k_{y1},\sigma_2}^{\dagger} \cdot \vec{D}_{n_3,k_{y1}+K_y,\sigma_3}\right]
  [n_F(n_2,k_{y1})  - n_F (n_3, k_{y1} + K_y) ] a_{n3,n_2,k_{y1}}(K_y) ,
\end{equation*}
and
\begin{equation*}
%\label{eq: xi m1m2k q}
  \xi_{m_1,m_2,k_y}(K_y) \equiv \sum_{\sigma_1,\sigma_2} \left[ \vec{D}_{m_1,k_y+K_y, {\sigma}_1}^{\dagger} \cdot \vec{D}_{m_2,k_y, {\sigma}_2}\right] \chi_{m_1,m_2}^{\sigma_1,\sigma_2}(K_y),
\end{equation*}
one finds
\begin{align}
\label{eq: plasmon eq dielectric form}
\xi_{m_1',m_2',k_y'}(K_y)  &= u_0  \sum_{m_1,m_2,k_y} \Bigg[L_y \sum_{\sigma_1',\sigma_2',  {\sigma}_2,  {\sigma}_1}
f_{m_1',m_2,m_1,m_2'}^{\sigma_1', {\sigma}_2, {\sigma}_1,\sigma_2'}
\left[ \vec{D}_{m_1',k_y'+K_y, {\sigma}_1'}^{\dagger} \cdot \vec{D}_{m_2',k_y', {\sigma}_2'}\right]
\left[ \vec{D}_{m_2,k_{y}, {\sigma}_2}^{\dagger} \cdot \vec{D}_{m_1,k_{y}+K_y, {\sigma}_1}\right] \nonumber\\
& \times
\frac{[n_F(m_2,k_{y})  - n_F (m_1, k_{y} + K_y) ]}{\omega_{k_y} - \varepsilon(m_1,k_y+K_y) + \varepsilon(m_2,k_y) }  \Bigg] \xi_{m_1,m_2,k_y}(K_y).
\end{align}
The quantity $\xi$ may be understood as a dielectric function, with
the expression inside the square brackets of Eq. \ref{eq: plasmon eq dielectric form} representing a generalized polarization function $\Pi_{m_1,m_2,k_y;m_1',m_2',k_y'} (K_y, \omega)$.
Non-trivial solutions to this equation must obey
\begin{equation*}
%\label{eq: plasmon generalized eigen problem}
  \text{det}(\mathbb{I} - u_0 \Pi(K_y, \omega)) = 0.
\end{equation*}

\subsection{Intra Subband Solutions for Two Subbands}
We now show how this equation leads to multiple gapless plasmons.  As a simplest concrete example we consider a massless chiral fermion system (e.g., single valley of graphene) with two subbands, both of which are occupied in the ground state, and include only intra-subband excitations in the analysis.  In Eq. \ref{eq: plasmon eq dielectric form}
this entails retaining only pairs of indices satisfying  $m_1=m_2$ and  $m_1' = m_2'$. One then has
\begin{align}
\label{eq: intraband plasmon eq dielectric form}
\xi_{m_1',m_1',k_y'}(K_y)  &= u_0 L_y \sum_{m_1,k_y} \Bigg[ \sum_{\sigma_1',\sigma_2',  {\sigma}_2,  {\sigma}_1}
f_{m_1',m_1,m_1,m_1'}^{\sigma_1', {\sigma}_2, {\sigma}_1,\sigma_2'}
\left[ \vec{D}_{m_1',k_y'+K_y, {\sigma}_1'}^{\dagger} \cdot \vec{D}_{m_1',k_y', {\sigma}_2'}\right]
\left[ \vec{D}_{m_1,k_{y}, {\sigma}_2}^{\dagger} \cdot \vec{D}_{m_1,k_{y}+K_y, {\sigma}_1}\right] \nonumber\\
& \times
\frac{[n_F(m_1,k_{y})  - n_F (m_1, k_{y} + K_y) ]}{\omega_{k_y} - \varepsilon(m_1,k_y+K_y) + \varepsilon(m_1,k_y) }  \Bigg] \xi_{m_1,m_1,k_y}(K_y).
\end{align}
The relevant spinors entering $\vec{D}$ can be evaluated as
\begin{equation}%\label{eq: graphene vec A}
\vec{A}_{n,k_y}  = \frac{1}{\sqrt{8 LL_y}} \frac{1}{\sqrt{k(k+k_y)}}
\left(
  \begin{array}{c}
    k+ k_y + ik_x \\
   i(k+ k_y - ik_x) \\
  \end{array}
\right), \nonumber
\end{equation}
and
\begin{equation}%\label{eq: graphene vec B}
\vec{B}_{n,k_y}  = \frac{1}{\sqrt{8 LL_y}} \frac{1}{\sqrt{k(k+k_y)}}
\left(
  \begin{array}{c}
    -(k+ k_y - ik_x) \\
   -i(k+ k_y + ik_x) \\
  \end{array}
\right). \nonumber
\end{equation}
where $k_x \equiv k_x(n)$ and $k=\sqrt{k_x^2+k_y^2}$.  We next note that for small $K_y$, the $k_y$ momentum sums involve only very small intervals, so that one may set all the values of $k_y$ in the various $\vec{D}_{m,k_y,\sigma}$ and $\vec{D}^{\dag}_{m,k_y,\sigma}$ factors appearing in Eq. \ref{eq: intraband plasmon eq dielectric form} to their values where the subbands cross the Fermi energy, $k_y \rightarrow k_F(m)$, where $k_F(m)=\sqrt{\varepsilon_F^2-(\hbar v_F k_x(m))^2}/\hbar v_F$, with $\varepsilon_F$ the Fermi energy and $v_F$ the velocity of the chiral fermion.  After some algebra, one finds
%\begin{equation}\label{eq: intraband V kernel, graphene}
%  \sum_{\sigma_1',\sigma_2',  {\sigma}_2,  {\sigma}_1}
%f_{m_1',m_1,m_1,m_1'}^{\sigma_1', {\sigma}_2, {\sigma}_1,\sigma_2'}
%\left[ \vec{D}_{m_1',k_y'+K_y, {\sigma}_1'}^{\dagger} \cdot \vec{D}_{m_1',k_y', {\sigma}_2'}\right]
%\left[ \vec{D}_{m_1,k_{y}, {\sigma}_2}^{\dagger} \cdot \vec{D}_{m_1,k_{y}+K_y, {\sigma}_1}\right]
%  \approx L \left(\frac{1}{LL_y}\right)^2 + \frac{L}{2} \left(\frac{ k_F(m_1)}{LL_y %\varepsilon_F}\right)^2 \delta_{m_1,m_1'}.
%\end{equation}

\begin{equation*}
%\label{eq: intraband V kernel, graphene}
  \sum_{\sigma_1',\sigma_2',  {\sigma}_2,  {\sigma}_1}
f_{m_1',m_1,m_1,m_1'}^{\sigma_1', {\sigma}_2, {\sigma}_1,\sigma_2'}
\left[ \vec{D}_{m_1',k_y'+K_y, {\sigma}_1'}^{\dagger} \cdot \vec{D}_{m_1',k_y', {\sigma}_2'}\right]
\left[ \vec{D}_{m_1,k_{y}, {\sigma}_2}^{\dagger} \cdot \vec{D}_{m_1,k_{y}+K_y, {\sigma}_1}\right]
  \approx
  \frac{1}{LL_y^2}\left[1+{1 \over 2}\frac{[\hbar v_F k_F(m_1)]^2}{\varepsilon_F^2}\delta_{m_1,m_1^{\prime}} \right]
\end{equation*}
Specializing to the case of just two occupied subbands, using the notation $\xi_{m,m,k_F(m)} \rightarrow \xi_m$, one finds to linear order in $K_y$
\begin{align*}
%\label{eq: 2 intraband plasmon eq}
\left(
  \begin{array}{c}
    \xi_1(K_y) \\
    \xi_2(K_y) \\
  \end{array}
\right)
= v_0
\left(
  \begin{array}{cc}
    (1+\frac{k_F(1)^2}{2\varepsilon_F^2}) I_1(K_y,\omega) & I_2(K_y, \omega) \\
    I_1(K_y, \omega)  &  (1+\frac{k_F(2)^2}{2\varepsilon_F^2}) I_2(K_y,\omega)\\
  \end{array}
\right)
\left(
  \begin{array}{c}
    \xi_1(K_y) \\
    \xi_2(K_y) \\
  \end{array}
\right)
\end{align*}
where
$v_0 = \frac{u_0 \varepsilon_F}{K_y L}$, $v_F\hbar=1$, and
\begin{equation}
\label{eq: I_m(q,omega)}
I_m(K_y, \omega) = \frac{K_y}{\varepsilon_F}\int_{k_F(m)-K_y}^{k_F(m)} \frac{d k_y}{\omega_{k_y} - \varepsilon(m,k_y+K_y)+ \varepsilon(m, k_y)}.
\end{equation}
Non-trivial solutons to Eq. \ref{eq: I_m(q,omega)} exist when
\begin{equation}
\label{eq: 2 intraband det}
  (a_1 I_1(K_y,\omega)-v_0^{-1}) (a_2 I_2(K_y,\omega)-v_0^{-1})  - I_1(K_y,\omega)I_2(K_y,\omega)=0,
\end{equation}
where
$a_m = 1+k_F(m)^2/2\varepsilon_F^2.$

For small $K_y$, Eq. \ref{eq: I_m(q,omega)} can be evaluated directly.  Writing the (non-interacting) speed of a particle along the wire in an occupied subband $m$ at the Fermi energy as $v_m$, for small $K_y$ one finds
\begin{equation}
\label{eq:I_evaluated}
I_m=\frac{K_y^2/\varepsilon_F}{\omega-v_mK_y}.
\end{equation}
Direct substitution of Eq. \ref{eq:I_evaluated} into Eq. \ref{eq: 2 intraband det} generates a quadratic equation for $\omega$ in terms of $K_y$, with solutions
\begin{equation*}
\omega_{\pm}(K_y) = {1 \over 2} \left\{v_1+v_2+(a_1+a_2)\tilde{u}_0 \pm
\left[\left(v_2+a_1\tilde{u}_0 -v_1 - a_2\tilde{u}_0\right)^2 +4\tilde{u}_0^2 \right]^{1/2} \right\} K_y,
\end{equation*}
where $\tilde{u}_0 \equiv u_0/L$.  Thus we generate two non-degenerate collective modes with frequencies different from those of the non-interacting particle-hole excitations.

\section*{Appendix C: Degeneracy of Transverse Dipole Moments}

In Section IV-A, it was shown that under certain circumstances the transverse dipole moment $\mathcal{D}_x(K_y)$ can be the same for multiple plasmon modes at small $K_y$, even when the frequencies of these modes are quite different.  This behavior is explained by Eq. \ref{eq: dipole explicit intraband, explicit}, in which one may see that $\mathcal{D}_x(K_y)$ becomes independent of the details of plasmon wavefunction when $m \gg \hbar v_F k_x(n)$ for the subbands $n$ involved in the plasmon wavefunction. (Note in Eq. \ref{eq: dipole explicit intraband, explicit}, $\hbar$ and $v_F$ have been set to 1.)  In this Appendix we present some details of the derivation of Eq. \ref{eq: dipole explicit intraband, explicit}.

Our starting point is Eq. \ref{eq: dipole explicit}, and we consider only intrasubband particle-hole excitations in constructing the low energy plasmon wavefunctions.  This means the quantities we need to focus on have the form
\begin{align*}\label{eq: d_m}
\vec{d}_{m}(k_y) = \int x \vec\psi^{\tau,*}_{m,k_y}(\vec{r}) \cdot \vec\psi^{\tau}_{m,k_y}(\vec{r})  d^2r.
\end{align*}
Writing the plasmon wavefunctions, Eq. \ref{eq: ribbon state}, in the form
\begin{equation*}
\vec{\psi}_{\vec{k}}^{\tau}(\vec{r}) =
   \vec{A}_{m,k_y}
   e^{i k_x x + i k_y y}
  + \vec{B}_{m,k_y} e^{- i k_x x + i k_y y},
\end{equation*}
where $k_x \equiv k_x(m)$, one finds
\begin{align*}
%\label{eq: d_m, simplify}
\vec{d}_{m}(k_y) = \frac{L^2 L_y}{2} (|\vec{A}_{m,k_y}|^2 + |\vec{B}_{m,k_y}|^2)
+ 2L_y \text{Re}( \vec{A}_{m,k_y}^{\dagger} \cdot  \vec{B}_{m,k_y})
\int x e^{-2i k_x x} dx
\end{align*}
Reading off the forms of $\vec{A}_{m,k_y}$ and $\vec{B}_{m,k_y}$ from the wavefunctions in Section II,
one finds
$$
|\vec{A}_{m,k_y}|^2 = |\vec{B}_{m,k_y}|^2 = \frac{1}{2L_y} \frac{m^2+k_x^2}{L(m^2+k_x^2) + \lambda_0 \tau m},
$$
where we have set $\hbar=v_F=1$.  Remarkably, these combinations are independent of $k_y$;
because $\vec{\mathcal{D}}(K_y) $ involves the difference of $\vec{d}_{m}(k_y)$ at two different $k_y$ values, terms involving $|\vec{A}_{m,k_y}|^2$ and $|\vec{B}_{m,k_y}|^2 $ do not contribute to the dipole moment of the plasmon.  For small $K_y$, the transverse component of the dipole moment can now be written as $\mathcal{D}_x \equiv \sum_m \mathcal{D}_{x,m}$ where the sum is over occupied subbands, and
\begin{align*}
%\label{eq: dipole individual subband, simp01}
  \mathcal{D}_{x,m}(K_y)
& = K_y \partial_{k_y} \left.\left(
   2L_y\text{Re}( \vec{A}_{m,k_y}^{\dagger} \cdot  \vec{B}_{m,k_y}
\int x e^{-2i k_x x} dx ) \right)\right|_{k_y=k_F} + \mathcal{O}(K_y)^2.
\end{align*}

With some algebra one may show
\begin{align*}
%\label{eq: A dagger B}
 \vec{A}_{m,k_y}^{\dagger} \cdot  \vec{B}_{m,k_y}
 = - \frac{( m E + i \tau k_x k_y )(m-i \lambda_0 \tau k_x) }{2 L_y E \left( L (m^2+ k_x^2)+  \lambda_0 \tau m  \right)},
\end{align*}
and using this relation, after performing the integration one finds
\begin{align*}
%\label{eq: dipole before derivatives}
2L_y\text{Re}( \vec{A}_{m,k_y}^{\dagger} \cdot  \vec{B}_{m,k_y}
\int x e^{-2i k_x x} dx  )
= \frac{  m  \lambda_0 \tau    }{2 \left( L (m^2+ k_x^2)+  \lambda_0 \tau m  \right)}
+ \frac{ \tau  k_y (L m + \lambda_0 \tau )  }{2  E \left( L (m^2+ k_x^2)+  \lambda_0 \tau m  \right)}.
\end{align*}
The first term is independent of $k_y$ therefore does not contribute.  Using
$$
\partial_{k_y} \left.\left(
\frac{k_y}{E}
 \right)\right|_{k_y=k_F}=\frac{m^2+ k_x^2}{\varepsilon_F^3},
$$
we arrive at
\begin{align*}
%\label{eq: dipole individual subband, simp02}
  \mathcal{D}_{x,m}(K_y)
 = \frac{K_y}{2 \varepsilon_F^3} \frac{ \tau  (L m + \lambda_0 \tau )  }{  L +  \frac{\lambda_0 \tau m}{m^2+ k_x^2}  } + \mathcal{O}(K_y^2).
\end{align*}
Thus to linear order in $K_y$, the transverse plasmon dipole moment is
\begin{align*}
  \mathcal{D}_x(K_y)
& = \sum_{\tau} \frac{\tau K_y  (L m + \lambda_0 \tau ) }{2 \varepsilon_F^3}
\sum_{n}\sum_{ k_y,s}
\frac{ |a_{n,n,k_y,\tau,s}(K_y)|^2  }{  L +  \frac{\lambda_0 \tau m}{m^2+ k_x(n)^2}  },
\end{align*}
which is Eq. \ref{eq: dipole explicit intraband, explicit}.

\end{widetext}

\bibliography{plasmon5}

\end{document}